\documentclass[]{interact}

\usepackage{epstopdf}
\usepackage{subfigure}
\usepackage{url}

\usepackage{color,soul}
\usepackage{natbib}
\bibpunct[, ]{(}{)}{;}{a}{}{,}
\renewcommand\bibfont{\fontsize{10}{12}\selectfont}

\theoremstyle{plain}

\theoremstyle{definition}

\theoremstyle{remark}

\begin{document}

\articletype{ARTICLE TEMPLATE}

\title{Optimising observing strategies for monitoring animals using drone-mounted thermal infrared cameras.}

\author{
\name{Claire Burke\textsuperscript{a}\thanks{CONTACT C.B. Email: C.Burke@ljmu.ac.uk}, Maisie Rashman\textsuperscript{a}, Serge Wich\textsuperscript{b,c}, Andy Symons\textsuperscript{d}, Cobus Theron\textsuperscript{e} \& Steve Longmore\textsuperscript{a}}
\affil{\textsuperscript{a}Astrophysics Research Institute, Liverpool John Moorse University, Brownlow Hill, Liverpool, L3 5RF, UK; \\\textsuperscript{b}School of Natural Sciences and Psychology, Liverpool John Moores University, James Parsons Building, Byrom Street, L33AF Liverpool, UK;\\
\textsuperscript{c}Institute for Biodiversity and Ecosystem Dynamics, University of Amsterdam, Sciencepark 904, Amsterdam 1098, Netherlands.\\
\textsuperscript{d}Department of Computer Science, James Parsons Building, Byrom Street, Liverpool, L3 3AF.\\
\textsuperscript{e}Drylands Conservation Programme, Endangered Wildlife Trust, Private Bag X11, Modderfontein, Johannesburg, 1645, South Africa.
}}

\maketitle

\begin{abstract}
The proliferation of relatively affordable off-the-shelf drones offers great opportunities for wildlife monitoring and conservation. Similarly the recent reduction in cost of thermal infrared cameras also offers new promise in this field, as they have the advantage over conventional RGB cameras of being able to distinguish animals based on their body heat and being able to detect animals at night. However, the use of drone-mounted thermal infrared cameras comes with several technical challenges. In this paper we address some of these issues, namely thermal contrast problems due to heat from the ground, absorption and emission of thermal infrared radiation by the atmosphere, obscuration by vegetation, and optimizing the flying height of drones for a best balance between covering a large area and being able to accurately image and identify animals of interest. We demonstrate the application of these methods with a case study using field data, and make the first ever detection of the critically endangered riverine rabbit ({\it Bunolagus monticularis}) in thermal infrared data. We provide a web-tool so that the community can easily apply these techniques to other studies (\url{http://www.astro.ljmu.ac.uk/~aricburk/uav_calc/}).
\end{abstract}


\section{Introduction}

Given the current high rate of species decline all over the world \citep{Ceb15,Regnier15}, there has never been a more pressing need for large-scale conservation efforts. Effective conservation strategy requires an understanding of locations, behaviours and survival requirements of animals \citep[see][]{Buc01,Schwarz08,Fre12,McM14}. However, monitoring of even large animals, which should in principle be easiest to detect, is a complex challenge. 

In the past, most surveys of animal populations have been carried out on foot or by light aircraft, such as planes and helicopters. These methods are time consuming and expensive, and in the case of aerial surveys potentially life threatening and limited to higher flight altitudes than might be ideal for sighting animals \citep{Wich18}. The proliferation of affordable, easy to use drones (also known as unmanned areal vehicles, UAVs, unmanned aerial systems, remotely piloted aircraft), with cameras mounted on them, presents the opportunity for surveying large areas cheaply and quickly. Surveying with drone-mounted optical (RGB) cameras can be effective for spotting animals in open homogenous terrain where the animals are not camouflaged. However, this is not often the case. 
The effectiveness of this type of study was demonstrated by \cite{Hodgson18} using a known number of life-sized replica seabirds arranged as a replica colony in an open area where the birds show clear colour contrast with their surroundings. When the numbers of replica birds counted were compared between numbers derived from drone-based images and ground-based images, the drone-based counts were found to be significantly more accurate.
However, in the review by \cite{Hodgson16} \citep[see also][]{Mara15} it is shown that issues with distinguishing animals clearly from their surroundings can detract from the potential increase in accuracy of animal counts offered by drone-based RGB footage. \cite{Duffy17} also point out that using drones for conservation ecology comes with its own challenges such as local environmental conditions and terrain. 

The addition of thermal-infrared (TIR) cameras increases the potential of drone-based studies by examining light that is emitted directly from sources, rather than reflected as is at visible wavelengths. Provided an animal's surface temperature (skin or coat) is warmer than its surroundings, as is the case for many homeothermic endotherms \citep{McCafferty15}, it will appear as a bright object at thermal infrared wavelengths. Since thermal infrared cameras only rely on the body heat of the targets to observe, rather then reflected light (from e.g. the sun), TIR cameras can be used at all times of the day or night, unlike RGB cameras which are essentially blind at night. This is especially advantageous for observing nocturnal species and has potential for use in anti-poaching efforts.

Handheld TIR cameras have been shown to be more effective for detecting simulated poachers on the ground at night than searching with a torch \citep{Hart15}. However foot surveys for poachers have obvious limitations in terms of the ground it is possible to cover, and are generally only used if poaching activity is suspected in the immediate area. It also requires a human operator to observe the thermal output, meaning scope for automation is limited. The advantage of drones to cover a large area quickly potentially allows for detecting poachers in both the immediate area of a ground patrol and further afield. Indeed \cite{Mara14} showed that drone-mounted TIR cameras are effective for seeing humans in the bush from the air.
The subsequent development of automated tools for spotting poachers with thermal-equipped drones, such at SPOT \citep{Bondi18}, demonstrates the power of this technology for making a major impact on poaching activity - both for prevention and as a deterrent. 
However, even with the rapid advancement of this technology there are many challenges to its implementation. In our previous paper \citep{Burke18a} we simulated poachers in the bush and attempted to quantify accuracy of detections from an automated system using observations from a drone-mounted TIR camera. Whilst the poachers could be detected, we find several sources of confusion from the environment, such as terrain and vegetation, and outline the steps which might be needed to address these. \cite{Mara14} also report difficulty with vegetation and terrain when attempting to detect humans on the ground with drone-based thermal hardware, and \cite{Bondi18} discuss potential difficulties with sufficient camera resolution for accurate detection at typical drone-flying heights.

With the recent increased availability of these technologies, many groups have also been looking to apply them to conservation and monitoring studies. However, as yet there is no universally accepted approach for best practice in their use.
Aerial surveys of populations of animals using TIR has been attempted by \cite{Burn09}, \cite{Witczuk17} and \cite{Mara14}, however these groups report that analysing the large volume of data produced and counting and identifying animals by eye is a very time-consuming method of data analysis.
\cite{Gooday18} also report several sources of confusion when it comes to reliably detecting and identifying animals with TIR, both drone-mounted and when handheld on the ground. With environmental features such as rocks changing temperature throughout the day acting as spurious sources, vegetation cover obscuring animals of interest, and the temperature itself acting as a source of confusion.

Since TIR data contains information about the temperature of the object being observed, it is possible to construct automated detection systems for objects with temperature distinct from their surroundings (warmer or colder). Automated detection has the additional advantage of removing the observer effect, meaning that the accuracy and reliability of wildlife counts can be more systematically quantified than detection performed by human observers. Different surfaces having different thermal properties also presents possibilities for automatically differentiating between spurious sources and targets of interest. Some attempts at automated detection of animals have been made in previous work \citep{Gonzalez15, Seymour17, Christiansen14, Isreal11, Longmore17, Chretien15, Chretien16, Israel17}, but these studies faced significant challenges in reliably detecting warm objects in the field, and separating animals of interest from spurious objects. Similarly, \cite{Lhoest15} construct a TIR-based algorithm to automatically count animals in groups. However, they find that their algorithm is not more accurate than counts by eye by a human analyst. An improved understanding of the data and how the environment effects the temperature values recorded will allow more reliable detection of animals and separation of spurious sources.
 All these groups have discussed significant challenges in terms of determining the optimal height and time of day to fly the drone in order to get high signal-to-noise detections, and subsequently reliable counts and identifications of the animals of interest. Flight strategy, and particularly flight height, may also be a consideration for minimising the disturbance of animals being observed. As discussed by \cite{Mara17}, terrestrial mammals and birds might respond to the presence of drones in various ways, and the response differs between species and individuals.

Within this paper we aim to address some of the issues encountered by previous studies in optimising drone flying strategy and understanding the effects of environment on TIR data. We also hope that this paper will function as a quick-start guide to those new to the field of drone-TIR observing, by compiling the fundamentals of use of this technology to one paper.

\subsection{The physics of thermal infrared observation}
All objects which have temperature emit energy in the form or electromagnetic radiation (light). The intensity of light emitted by an object at a given wavelength depends on the temperature of the object. With typical body surface temperatures of $\sim20-30^{\circ}$C, humans and other homeothermic animals emit most strongly in the wavelength range between 8--14 $\mu$m (8$\times 10^{-6}$--14$\times10^{-6}$ meters). That makes this region of the electromagnetic spectrum, known as the thermal infrared, an ideal wavelength with which to detect animals. The luminosity, $L$, of this light varies as a function of temperature, $T$, by,
\begin{equation}
L=A\sigma T^4,
\label{equ:LT}
\end{equation}
where $\sigma$ is the Stefan-Boltzman constant, $A$ is the emitting area (meters$^2$), and all values of temperature are in units of Kelvin (0$^{\circ}$C=273K). As such the luminosity (brightness) of an object increases strongly as its temperature increases. 

Thermal infrared cameras are sensitive to the luminosity of objects in the TIR wavelength range, and allow this luminosity to be converted in to images which are intelligible to humans. In reality, the luminosity recorded by TIR cameras can be affected by several factors, these include the properties of the camera itself and the properties of the environment in which observations are made. As discussed in the reviews by \cite{cilulko} and \cite{HavensSharp} these issues include weather conditions, temperature contrast between object and background, distance between object of interest and camera (affects of atmosphere and accuracy of camera), field of view, effects of obscuration by vegetation, physical properties of animal coat, animal behaviour \citep[see also the difficulties reported by][]{Gooday18}. 

In order to accurately interpret infrared data, to be able to reliably detect and identify different animals, it is important to know how well the TIR camera used to take the data performs in different environments. It is also necessary to have an understanding of how different environments and different ways of using the camera will effect what can and can't be seen or what features can be distinguished.
In this paper we apply physical understanding to provide potential solutions to the issues related to observing environment - namely weather, background emission, distance to object and field of view of the camera, absorption by the atmosphere, obscuration by vegetation. We combine these to create a recipe for optimum strategies for observing animals in the wild based on their temperature and the environmental conditions in which they are observed. 
We also publish an accompanying web-tool so that future users can easily calculate and account for these effects in their observations.

This paper is split into two parts - in sections~\ref{sec:fov}--\ref{sec:veg_mixing} we discuss some of the general challenges faced when observing using TIR cameras, and present some steps that can be taken to resolve these. In section~\ref{s:case_study} we present a case study in which we show how the methods we describe can be applied to overcome observing challenges, and in section~\ref{lit_comparson} we discuss how these methods could have been applied to previous studies to optimize observing strategies.

\section{Field of view and drone height}\label{sec:fov}
In order to acquire TIR data of a homeothermic animal, which can be used for scientific or conservation purposes, it is important to consider the size, shape, body temperature and habitat of the animal. To accurately identify a species of animal in the footage it must appear larger than some minimum area (in pixels) in the field of view of the camera. This is true for both TIR and RGB cameras. Additionally most TIR cameras will not accurately record the temperature of an object if it is smaller than a certain minimum area in pixels -- the animal must be well resolved (this will be discussed more fully in Section~\ref{sec:veg_mixing}). Drone surveys must balance covering the maximum area on the ground in the time available to fly with having the necessary minimum resolution for animals to be resolved and identified. The projected area viewed on the ground and size of the animals within the field of view depend on the flying hight of the drone.

Every camera has an intrinsic angular pixel scale, which is defined as the angular size in degrees covered by each pixel. This is given by,
\begin{equation}
\frac{\theta}{\#pixels}= \rho_{a},
\label{equ:ang_pix_scale}
\end{equation}
where $\rho_{a}$ is the angular pixel scale, $\theta$ is the field of view (FOV) in degrees, and $\#pixels$ is the number of pixels in the detector. TIR cameras generally have square pixels, so the angular pixel scale will be approximately the same along both horizontal ($x$) and vertical ($y$) axes.
If a camera is mounted on a drone so that it points straight down towards the ground (the nadir), the projected field of view will be a fairly undistorted rectangle or square depending on the number of pixels along the $x$ and $y$ axes of the detector in the camera. As such we can define a physical pixel scale,  $\rho_{p}$, of the camera -- that is the length in the $x$ and $y$ directions covered by each pixel in meters on the ground -- as a function of height, h, by; 
\begin{equation}
h~tan(\rho_{a})= \rho_{p}.
\label{equ:pix_scale}
\end{equation}
The projected field of view of the camera on the ground in meters is then simply the physical pixel scale multiplied by the number of pixels in the detector. Using this relation we can calculate the height for a given camera to attain a minimum target area, length or width in the field of view, and the area covered in the field of view at this height. 

For example, the minimum diameter for accurate identification of an object (or species of animal) and measurement of its temperature is 10 pixels for most TIR cameras (see Section~\ref{sec:veg_mixing} and Figure~\ref{spot_size} for a full discussion). To detect an animal 1~meter in length when viewed from directly above, given a FLIR Tau 640 model TIR camera with a 13 mm lens, which has $\#pixels = 640 \times 512$ pixels and $\theta = 45^{\circ}\times37^{\circ}$, for a pixel scale of  $\rho_{p}=0.1$ meters/pixel (10~cm per pixel, which gives a length of 10 pixels in total for a 1 meter animal) then the  drone should fly at $h=81.5$ meters above ground level (AGL), and the field of view on the ground will be 67.5 $\times$ 54.5 meters. This calculation can be performed simply for any camera using our observing strategy optimization web-tool (\url{http://www.astro.ljmu.ac.uk/~aricburk/uav_calc/}).

\begin{figure}
\includegraphics[width=8cm]{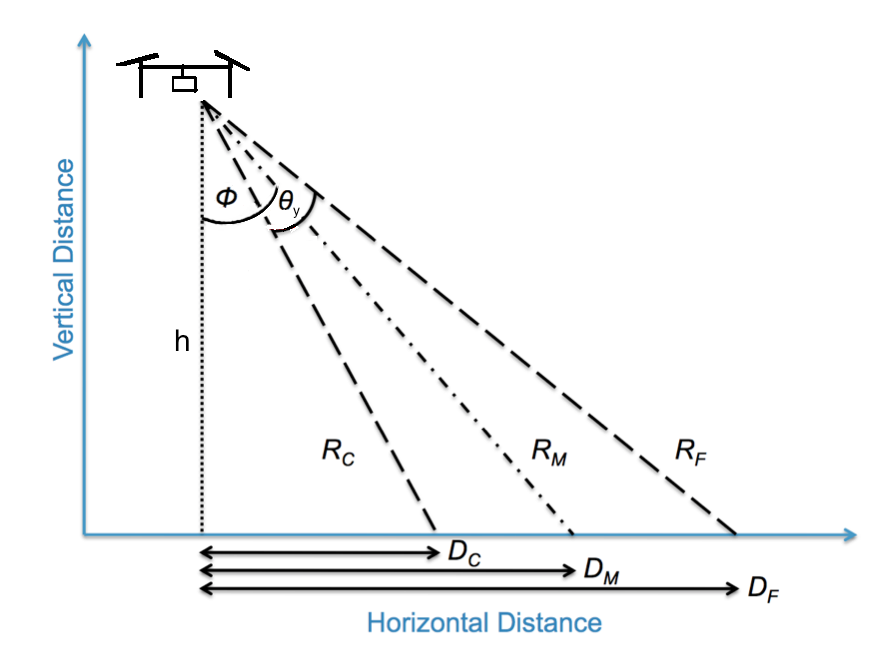}
\includegraphics[width=6cm]{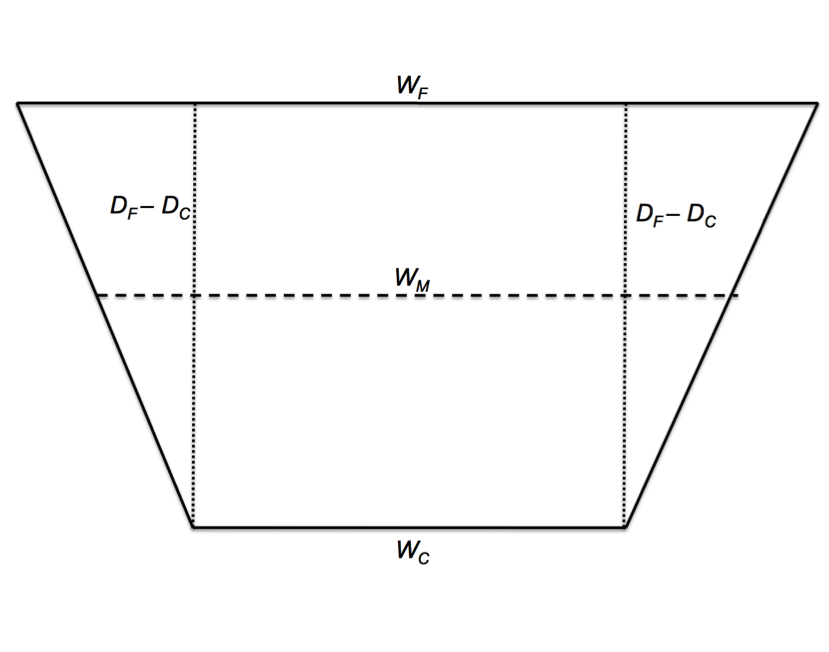}
\caption{Left: Angled camera set up on drone, $\phi$ is the angle of the camera with respect to the ground, $\theta_y$ is the angular FOV of the camera along its y-axis, $D_C, D_M, D_F$ are the closest, middle and furthest distances along the field of view of the camera, $h$ is the drone height. $R_C,~R_M,~R_F$ are the distances between the drone and the bottom, middle and top of the field of view. Right: The de-projected field of view showing the area on the ground covered by the camera resulting from this setup. $W_C, W_M, W_F$ are the widths of the field of view on the $x$-axis of the camera for the closest, middle and furthest distances along the $y$-axis. \citep[Adapted from ][]{Longmore17}.}
\label{fov_fig}
\end{figure}

Alternatively, a TIR camera may be mounted on a drone at an inclined angle to the ground, see Figure~\ref{fov_fig}. This strategy may be effective for gathering data on smaller animals which hide beneath vegetation, or for covering a larger instantaneous area on the ground with the camera's field of view. This can also help with data interpretation by human eye as the targets will stay in the FOV for longer. With an angled setup the field of view on the ground is distorted -- covering a wider area on the ground at the top of the field of view and a narrower area at the bottom, see Figure~\ref{fov_fig}, therefore every pixel will cover a differently sized physical area on the ground and have a different physical pixel scale. 
The calculation of individual pixel scales for all pixels in the FOV is discussed in Appendix~\ref{app:angle_math}. 
Assuming the desired pixel scale is in the centre of the field of view for an angled setup, the value for height calculated becomes $R_M$ in Figure~\ref{fov_fig}, corresponding to the distance between the camera and the centre of the field of view on the ground. The relation between flying height of the drone is then given by,
\begin{equation}
h=R_M.Cos\left(\phi-\frac{\theta_y}{2}\right).
\label{equ:any_R}
\end{equation}
For example, using the same target and camera as above, if the camera is instead mounted at a $\phi=60^{\circ}$ angle, in order for the target to be 10 pixels in diameter in the $middle$ of the field of view, then $R_M =81.5$ meters would be needed, giving $h=40.75$ meters AGL. Since the apparent size of objects will vary across the field of view of a camera with the angled setup this could potentially make automated detection tricky, and care should also be taken when attempting to identify sources by eye in these cases.

\section{Absorption by intervening medium}\label{sec:absorb}
\begin{figure}
\includegraphics[width=14cm]{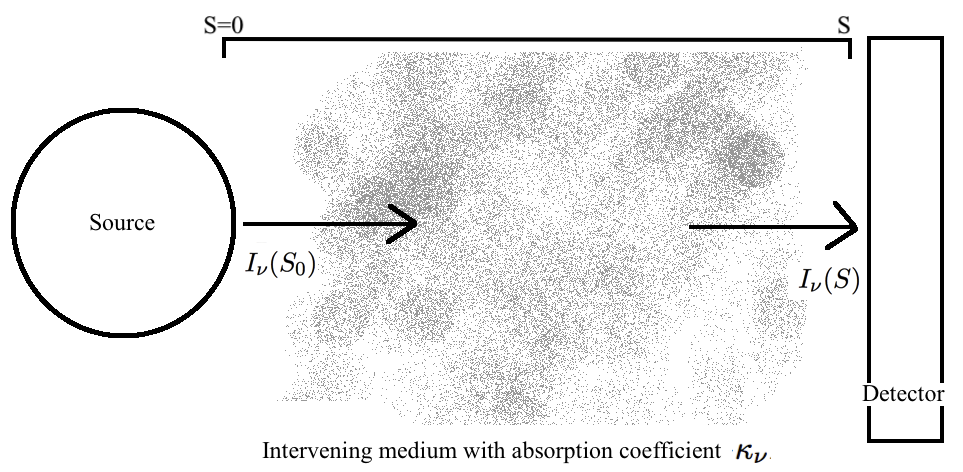}
\caption{Figure illustrating absorbing medium between source and detector. $I_{\nu0}$ is the intensity of thermal infrared light emitted by the source,  $I_{\nu}$ is the intensity of TIR light received by the detector as a result of it traveling a distance $S$ through an intervening medium with absorption coefficient $\kappa_{\nu}$.}
\label{abs_coeff}
\end{figure}

In order to perform accurate measurements of the temperature of objects or animals being observed it is necessary to have a thorough understanding of the observing conditions. The properties of the medium that emitted radiation passes through before reaching the detector (camera) will have an effect on the signal received, illustrated in Figure~\ref{abs_coeff}. The change in intensity of radiation as it passes through a given medium is dictated by the physics of radiative transfer, which is described in Appendix~\ref{app:rad_tran}.

TIR radiation that is emitted by animals or humans must pass through some amount of the Earth's atmosphere before it is detected. Molecules in the atmosphere absorb and emit radiation. For example, water vapor is known to be a particularly strong absorber of radiation at thermal infrared wavelengths \citep{GORDON07}. The amount of absorption and emission of radiation by molecules in the atmosphere depends on the pressure, temperature and relative abundance of different species of molecule \citep{hitran_b}. The major constituents of the atmosphere in the lower troposphere are nitrogen, oxygen and water. The relative abundance of these molecules is largely described by humidity, thus the amount of TIR radiation from as source which is absorbed by the atmosphere is a function of temperature, pressure and humidity. 

To determine the effect of atmospheric absorption on the TIR measurement made with thermal cameras on drones we take data from the HITRAN molecular spectroscopic database \citep[][for full details]{hitran, hitran_b} for a mixture of gases representative of Earth's atmosphere at various values of humidity, pressure and temperature. From this an absorption spectra can be generated (described in Appendix~\ref{app:rad_tran}), which shows how much radiation from a source is absorbed as a function of its wavelength. Using the physics of radiative transfer (App~\ref{app:rad_tran}) one can then derive the fraction of light absorbed and re-emitted by the atmosphere as a function of distance between source and observer for a range of values of temperature, pressure and humidity, and convert this into a change in observed temperature.

\begin{figure}
\includegraphics[width=7cm]{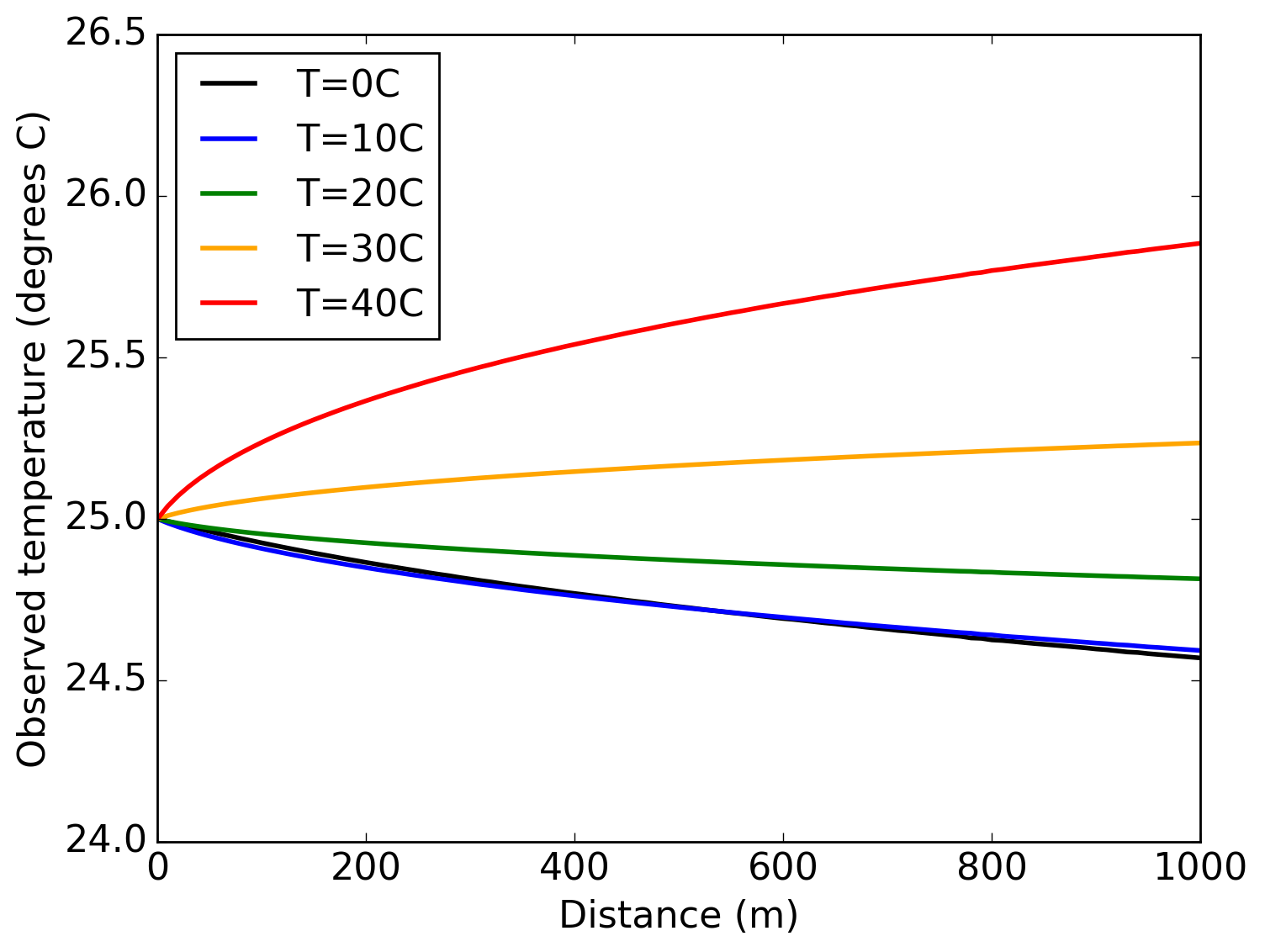}
\includegraphics[width=7cm]{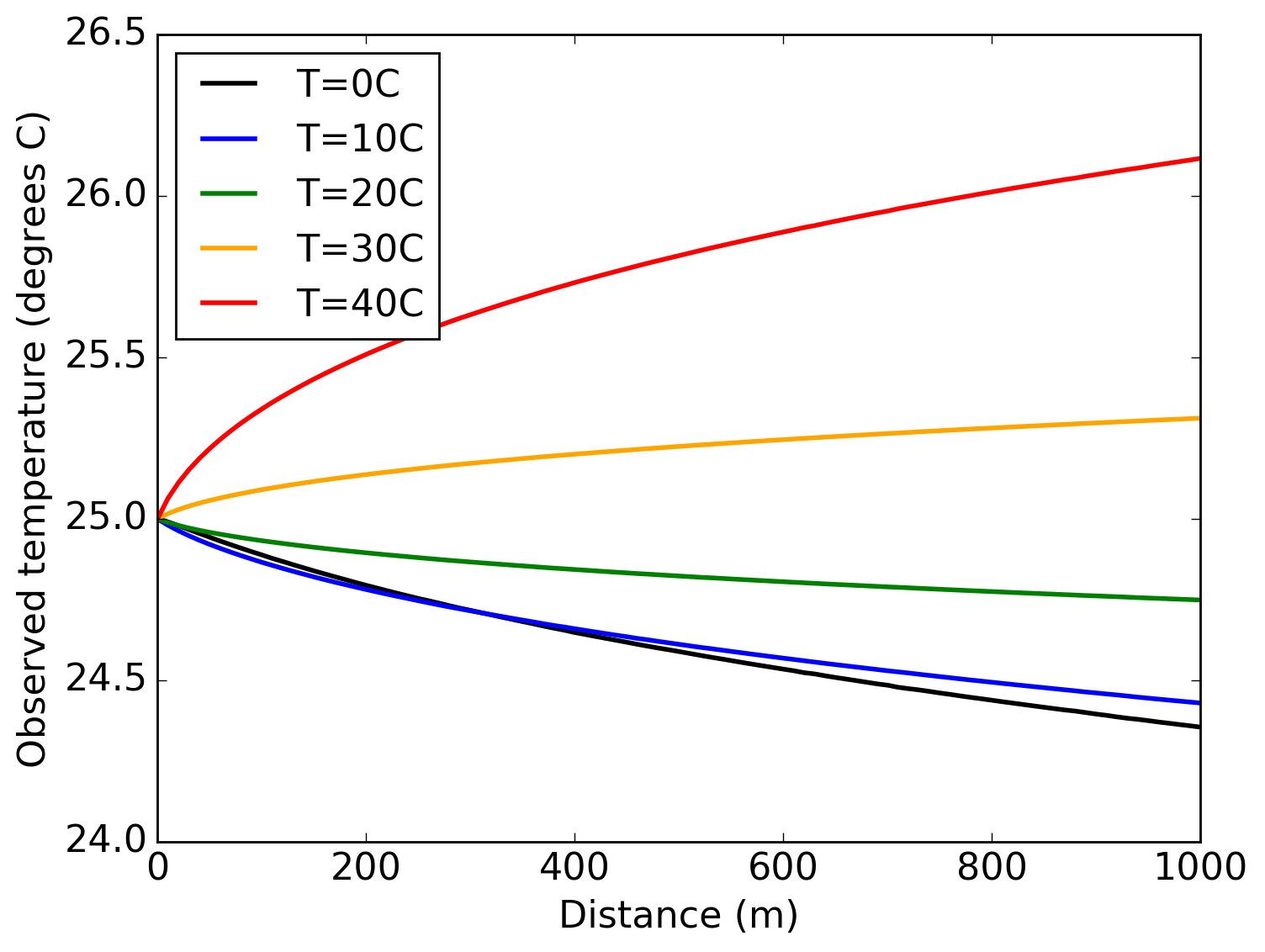}
\caption{
The temperature that would be recorded for an object with actual temperature $T_{s}=25^{\circ}$C when viewed under different atmospheric conditions,  calculated using Eq~\ref{equ:dT} described in Appendix~\ref{app:rad_tran}.
Left: The temperature that would be recorded, $T_{obs}$, vs distance from source as a result of the absorbed and emitted radiation by the atmosphere with air temperature $T_m=0-40^{\circ}$C ($\equiv$ 273--313 K), pressure 101 kPa ($\sim$1 standard atmosphere of pressure) and 50\% humidity.
Right: The temperature ($T_{obs}$) that would be a recorded in a more extreme atmospheric scenario (similar to that which might be found on a hot day in a tropical rain forest) of 110 kPa pressure and 100\% humidity for a range of values of $T_m$.}
\label{fig:spectr_abosorb}
\end{figure}

Figure~\ref{fig:spectr_abosorb} shows the observed change in temperature for an object with actual temperature $T_s=25^{\circ}$C, when viewed through atmospheric air with a range of temperatures $T_m$, and typical values of pressure and humidity. For TIR cameras with typical sensitivity $0.05^{\circ}$C this change will be observable. The effect of the temperature of the air is clear in the figure, when the atmospheric air temperature is lower than the source temperature, the observed temperature is up to half a degree lower than the actual source temperature (within the first 100 meters from the source). However if the air temperature is warmer than the source temperature, a source will appear to be warmer than it actually is. The change in source temperature observed depends on the difference between source and air temperature, so will need to be calculated for different observing conditions and targets. This will be made possible to calculate through the web-tool.

Since amount of atmospheric absorption changes as a function of temperature, pressure and humidity, for higher pressure and humidity the effect of the atmosphere will be stronger. The right hand panel of Figure~\ref{fig:spectr_abosorb} show an extreme case of 110kPa pressure and 100\% humidity -- conditions which may be found in a tropical rainforest. 

In many cases for off-the-shelf TIR cameras, the $absolute$ measurement uncertainty of the camera is $\pm5^{\circ}$C. However, the $relative$ temperature sensitivity (the ability to measure small changes in temperature differences between pixels) is typically much more accurate, on the order of 0.05$^{\circ}$C. So whilst the effect of absorption and emission of TIR radiation by the atmosphere can be accounted for using the atmospheric absorption spectra, this affect will be increasingly important for detection of objects at larger distances, i.e., when flying higher than 100 meters AGL. This is also an issue when observing using satellites. However at distances 1km or greater the temperature, pressure and humidity of the atmosphere will not be uniform across the whole distance and more sophisticated atmospheric modelling is required. 

For drone monitoring purposes, the effect of the atmosphere may be particularly noticeable when using an angled camera setup, as the distance between the sources and camera at the top of the field of view can be much greater than at the bottom for large angles of $\phi$ (see Figure~\ref{fov_fig}). Sources at the top of the field of view will appear warmer or cooler in the data due to them being observed through more of the atmosphere. In this case the temperature change due to the atmospheric conditions should be calculated for the distances ($R$ in Figure~\ref{fov_fig}) at both the top and bottom of the field of view. An example of this, and the effect of viewing objects through large amounts of atmosphere with the angled camera setup is given in Appendix~\ref{app:rad_tran}.

In summary, the ambient atmospheric conditions affect how much TIR radiation from sources is received by cameras. This effect is small for typical drone flying heights, if the camera is pointed straight down but can become a significant source of error for very precise measurements, flying at larger heights, or if the camera is at a large angle from nadir so viewing the ground at a larger range of distances. Figure~\ref{corrections_arrays} illustrates how the atmospheric absorption and emission and angle that the camera is mounted at will affect observations.

\subsection{The effect of fog}
\begin{figure}
\includegraphics[width=7.5cm]{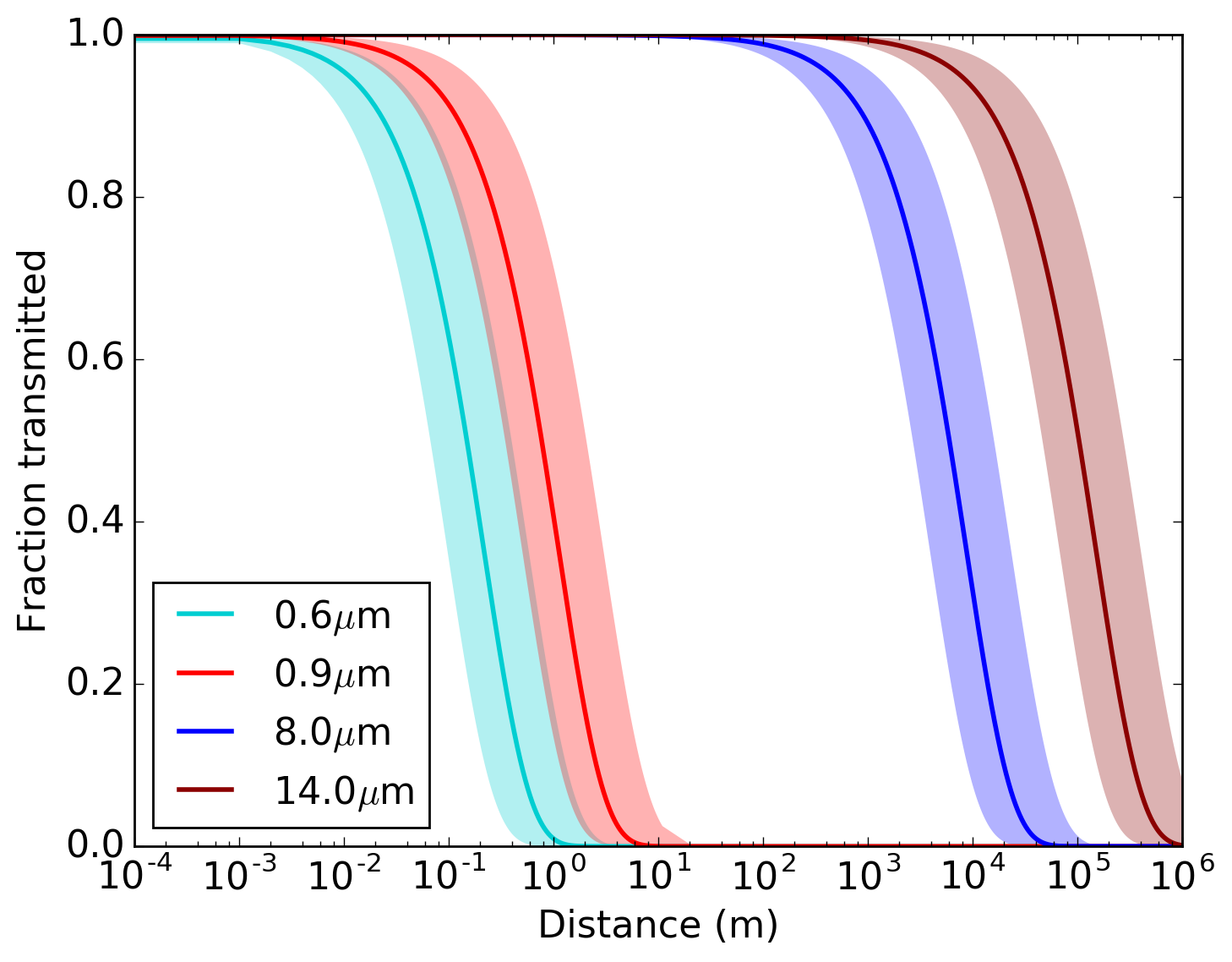}
\includegraphics[width=7.5cm]{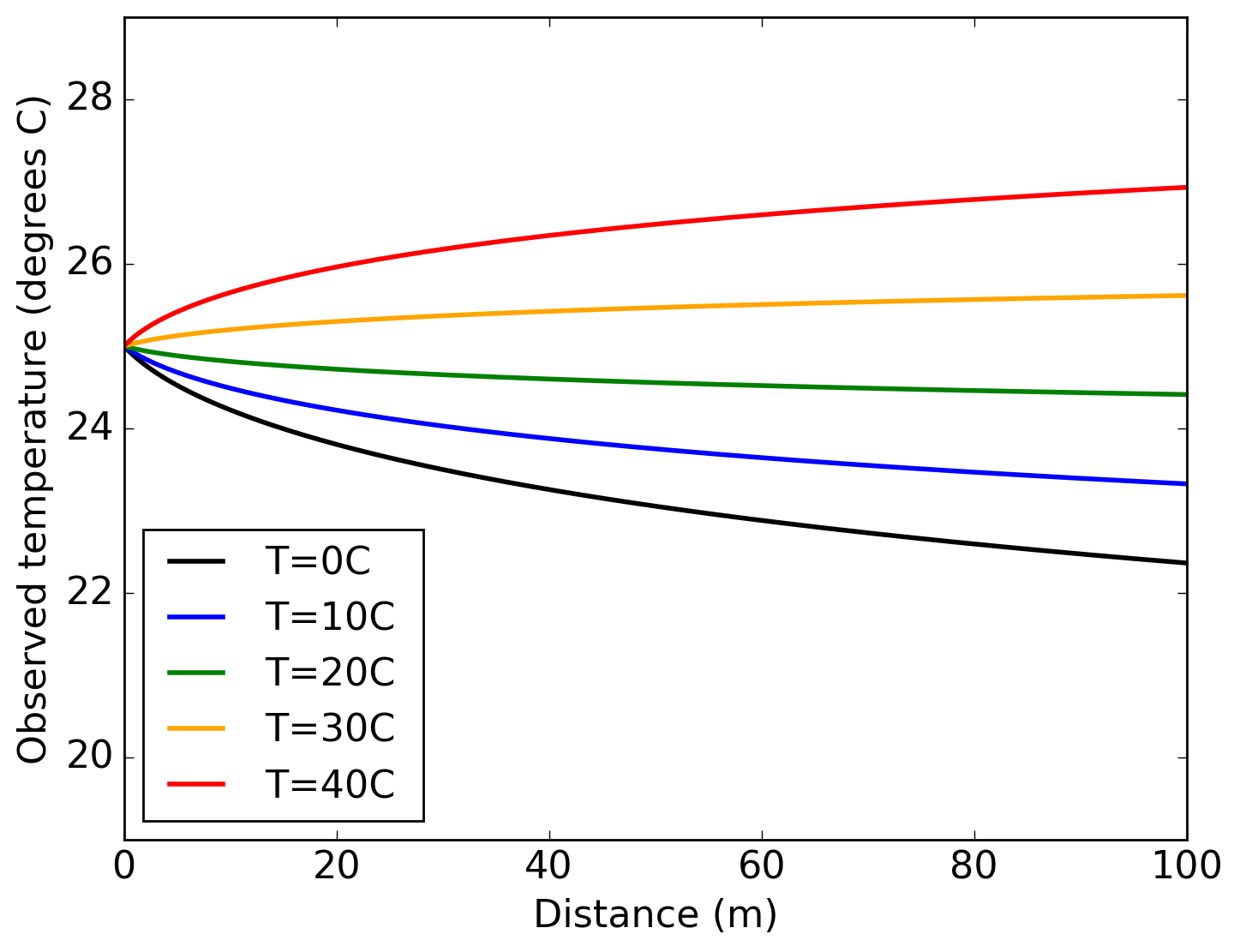}
\caption{Left: Fraction of light from sources transmitted when passing through air containing suspended fog droplets at different wavelengths, $0.6-0.9\mu$m covers the visible range of light, $8-14\mu$m is the range covered by thermal infrared light. Right: Change in observed temperature of object with actual temperature $T_s=25^{\circ}$C as a result of absorption and emission by fog for a range of air temperatures $T_m=0-40^{\circ}$C, with pressure P=110 kPa. The absorption spectra for water vapour is largely unaffected by the range of pressures experienced within the Earth's atmosphere. Humidity is always H=100\% for fog. }
\label{fog_fig}
\end{figure}

Fog is a major issue for flying and observing with any aircraft. Fog occurs when the air is already saturated with water vapour (humidity is 100\%) but evaporation is still occurring. An example of conditions which may cause this is if the ground or a body of water is warm enough for evaporation, but the air above is very cold and therefore already holding as much moisture as it physically can between other gaseous molecules. The suspended droplets of water in fog have typical diameters of $1-10\mu$m and the number density of droplets is between $1-100$ cm$^{-3}$ \citep{fog_ref, fog_ref2}, with larger droplets having a smaller number density per unit volume. As visible light passes through the fog it is scattered by these droplets \citep{scattering_book}, however due to its longer wavelength TIR light is scattered less. The result of this is that much more TIR light can pass through and it is possible to see objects much further away then with visible light. 
Figure~\ref{fog_fig} shows how light of visible and thermal infrared wavelengths is scattered by thick fog \citep[constructed using data taken from][]{water_ref_index, fog_ref, fog_ref2}. As is clear in this figure, for visible light 50\% of the emitted radiation from the source is transmitted through the fog up to a distance $\sim1$ meter, the rest is scattered. For TIR light the distance beyond which 50\% of light is scattered is 10,000 meters.
 See Appendix~\ref{fog_appendix} for a detailed physical description of the effect of fog on light.

We generate an atmospheric spectra for water vapour in the same manner as that described above for other atmospheric gas mixtures. An example of the subsequent change in temperature of an object that would be observed as a result of fog is shown in Figure~\ref{fog_fig} for a range of different air temperatures. For the range of pressures experienced within the Earth's atmosphere the absorption spectra for water vapour is largely constant, so temperature is the only variable which needs to be considered for absorption by fog. In order to account for the effect of fog on observations, read the change in temperature for the height of the fog, then add this to the change in temperature for the height of the drone as a result of the normal atmospheric gases. For example, for an object with actual temperature of $T_s=25^{\circ}$C, if using a drone-mounted TIR camera in the conditions described in the bottom row of Figure~\ref{fig:spectr_abosorb}, namely air temperature $T_m=40^{\circ}$C, pressure P$=110$kPa and humidity H$=100$\%. If the depth of the fog is 20 meters (from the ground to the top of the fog) and the drone height $h$=100 meters, then the observed change in temperature due to the fog will be $+1^{\circ}$C (see Fig~\ref{fog_fig}) and the observed temperature change due to the rest of the gases in the atmosphere will be $+0.5^{\circ}$C (Fig~\ref{fig:spectr_abosorb}), giving a total change in observed temperature for objects observed of $+1.5^{\circ}$C ($T_{obs}=26.5^{\circ}$C).

Whilst we have shown that fog itself does not affect the transmission of TIR radiation significantly, even for the case of very dense fog shown here, it is  still the case that cold moist air can have an effect on TIR cameras. For example if a TIR camera is mounted to a drone at an angle, and that angle is pointing forwards with the direction of flight of the drone, there is an increased risk of condensation forming on the lens due to the high moisture content of the air. TIR radiation cannot pass far through condensed bodies of water, so condensation on the lens of the camera will cause a significant reduction in data quality. The presence of larger condensed droplets of water in a column of fog may also reduce the quality of data recorded, however modelling of this is beyond the scope of this paper. Flying in cold moist air may also lead to the camera housing cooling significantly and this can affect the performance of the detector. 

\section{Emission from background sources}
\begin{figure}
\includegraphics[width=15cm]{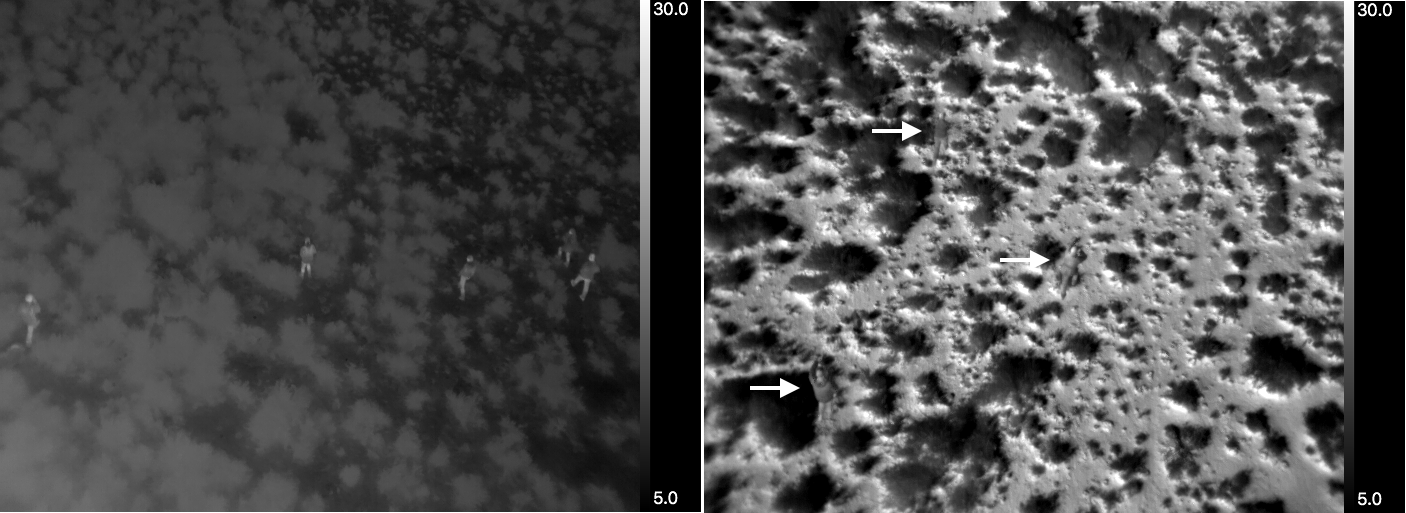}
\caption{Data from 2 flights over the same terrain on the same day (26/09/2017) at 2 different times with different LST. Drone height was $h=15$ meters, with the camera mounted at a $\phi=45^{\circ}$ angle. The field of view is $D_F-D_C=22.6$ meters, $W_C=$13.4 meters, $W_F=$28.3 meters . Left: Data taken at 0800, 5 humans are clearly visible. Right: Data taken at 0930 showing 3 humans indicated with arrows. Scale on the right hand side of both panels is the temperature in $^{\circ}$C.}
\label{fig:lst_comparison}
\end{figure}

The major advantage of observing homeothermic species at thermal infrared wavelengths is that since they are often warmer than their surroundings, they appear brighter in TIR images. This property negates issues presented in visible wavelengths such as camouflage and some amount of obscuration by environment. However, if the temperature of the surroundings is similar to that of the animal being observed, then the brightness contrast between the desired target and the background will be low. In extreme cases, when the background is warmer than the animal, the primary advantage of TIR imaging is lost, making it difficult to distinguish the target from the background (see Figure~\ref{fig:lst_comparison} for example). Clearly it is critical for successful monitoring to know in advance the temperature properties of the animals to be observed and the typical background temperature of their environment, so that the time of day (etc) for drone missions can be optimized for animal detection and identification. We propose partial solutions to this issue using satellite data.

Satellites have been observing the Earth for several decades in order to monitor the weather and changes in climate. Variables monitored include the land surface temperature (LST). LST is a direct measure of the temperature of the surface (or ground) and is distinct from the air temperature (which we humans tend to feel when we go outside). LST is most strongly regulated by incoming solar radiation -- so LST is often at its peak when the sun is directly overhead for any given region of land. This means that the peak LST does not always coincide with the peak air temperature, commonly observed by people as the warmest part of the summer. LST observations from satellites provide a complementary dataset for thermal infrared drone-based observations, making it possible to forecast the best time of year to observe with TIR-mounted drones, and potentially making it possible to correct for (or subtract) the emission from the ground, thereby providing better quality data on objects of interest (animals, poachers, etc).

We use land surface temperature data from the Moderate Resolution Imaging Spectroradiometer (MODIS) satellites. MODIS data comes from 2 satellites on Sun-synchronous near-polar orbits, which collectively pass overhead at each point on Earth's surface 4 times a day (0130, 1030, 1330, 2230 local solar time). The data was verified with in-situ measurements and found to be accurate to within 1$^{\circ}$C \citep[][for full details]{modis_ver}. We used an LST climatology generated from MODIS data for the years between 2003-2014 \citep[see][]{modis_clim}. Averaging over the whole 11 year period of the observations allows the variations in weather and cloud cover on individual days to be accounted for. The climatology has a resolution of 1km$^2$ covering all the land surface on the globe. The 1~km$^2$ resolution is fine enough spatial scale for resolving different terrains (mountains vs planes, etc), but cannot resolve individual ground features (rocks, trees, etc). Given regular surveying of an individual site with a drone-mounted camera it would be possible to construct a ground surface model and thus generate a more detailed model of how the ground heats up throughout the day (for that specific site),  however this type of survey can be very time consuming and reduces the cheap, fast advantage conveyed by using drones.

\begin{figure}
\centering
\includegraphics[width=7cm]{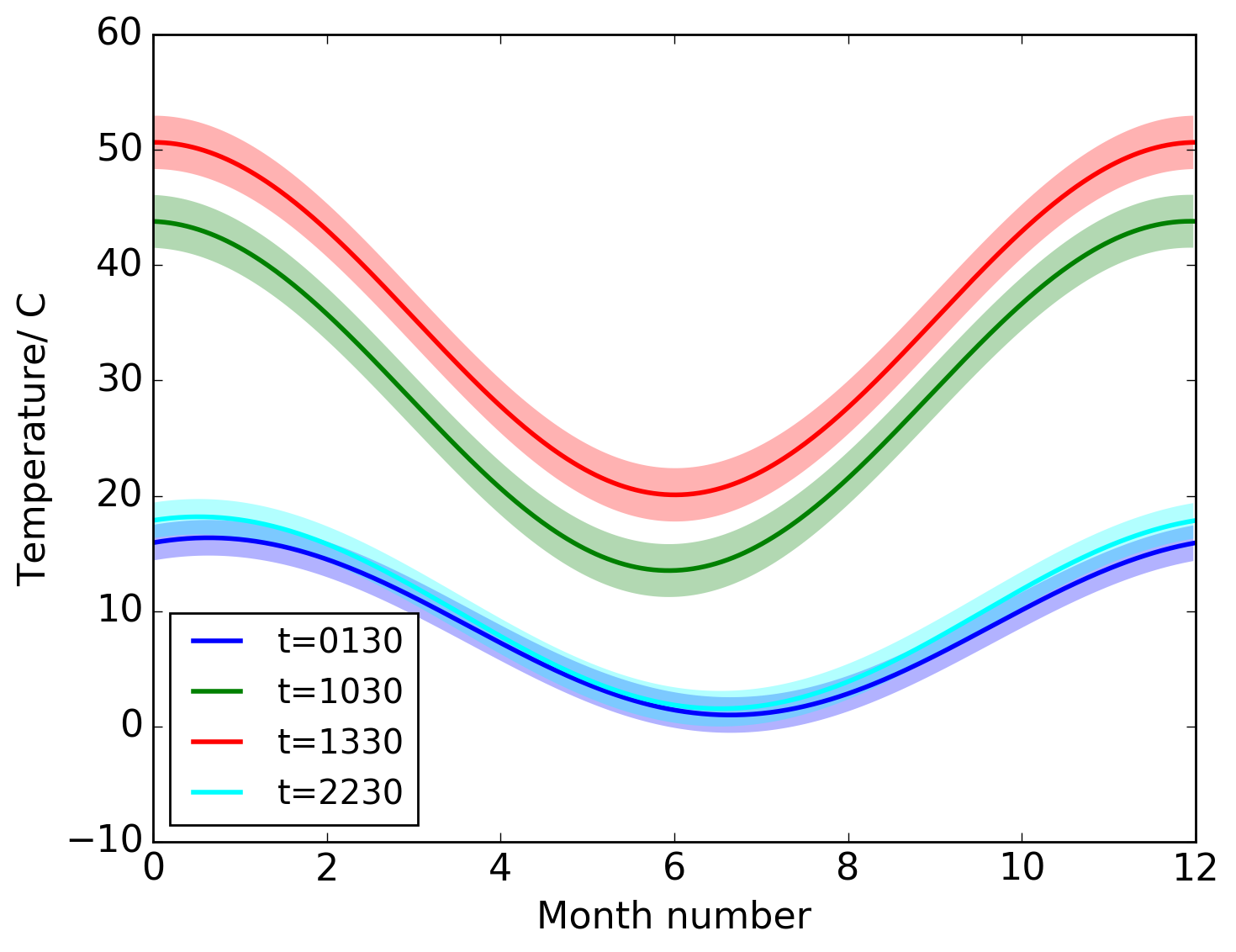}
\includegraphics[width=7cm]{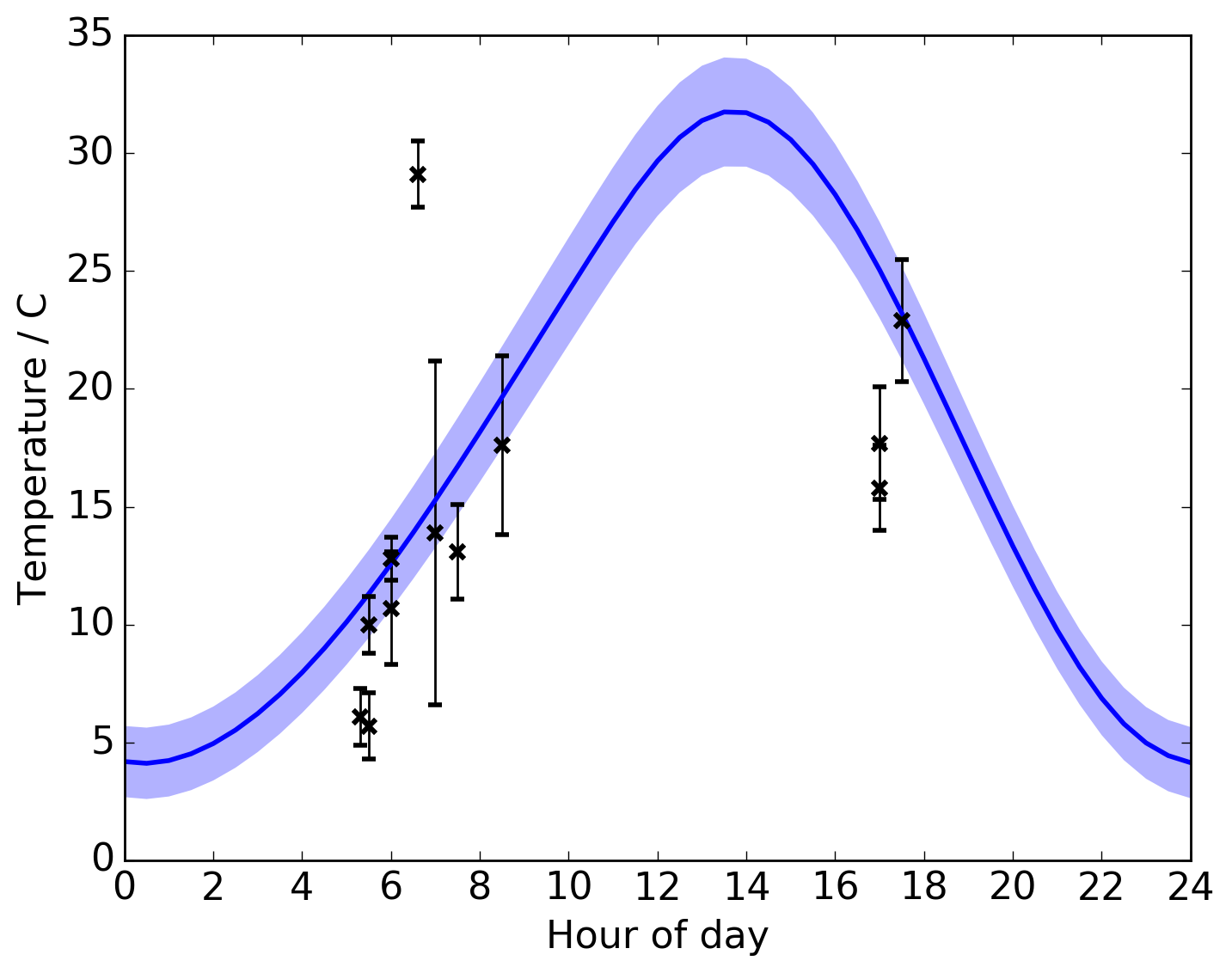}
\caption{\label{fig:loxton_lst}Example LST climatology for Loxton, South Africa in the 1 km$^2$ centred on the observations. Solid lines indicate the climatological mean, shading indicates the 2-sigma inter-annual variation. Left: annual LST climatology - different colours represent different times of day. Right: diurnal temperature range for the 20th September (climatological mean), `x' indicates the mean observed temperature from data taken between 21st--27th September 2017, the error bars indicate the 2-standard deviation from the mean range of temperatures recorded for the ground. }
\end{figure}

Figure~\ref{fig:loxton_lst} shows an example of the 11-year mean and 2-standard deviation (2$\sigma$) range of a 1 km$^2$ region in South Africa (see Section~\ref{s:case_study}). The seasonal cycle can be seen as an annual peak and trough in temperatures for all four times of day. It should be noted that the annual temperature cycle does not necessarily peak in the middle of summer, as the peak in incoming solar radiation is not necessarily at 21st June (Northern Hemisphere) or 21st December (Southern Hemisphere), but varies as a function of latitude depending on when in the year the sun is overhead at midday. Additionally, the ground retains some of the heat from the Sun's radiation for some time after it is received. How long it takes to lose this heat depends on the properties of the surface. The variation in peak of the annual temperature is particularly obvious in the tropics. In the tropics (and near the equator), the Sun is overhead at midday twice per year, this can lead to a double peak in annual LST or an offset between peak day and nighttime temperatures. Regions in the tropics will also show a much smaller seasonal variation than temperate regions as a result.

The internal and external (skin) temperature of an animal are rarely the same, especially in larger animals where core to surface temperature gradients can be prominent \citep{McCafferty15}. 
To detect an object of given (fixed) temperature it is best to observe during the coldest time of the year for LST in order to get high signal-to-noise detections. In reality, homeothermic animals can alter their skin temperature throughout the year so that they have a low temperature gradient with their surroundings, so as not to lose too much body heat. As a result of this, animal skin (or coat) temperatures may appear cooler during winter months. Even so, for higher signal-to-noise detections colder months and colder times of day are still preferable for observing as the difference in temperature between the source and the ground will still be larger as a fraction (or ratio) than in the warmer times.
By interpolating over the 4 different times of day (using a cubic spline) we can derive the daily diurnal temperature range for a given day of interest, see Figure \ref{fig:loxton_lst} for example. Using this it is possible to determine in a statistical sense the best time of day for any given time of year to observe. We test the LST climatology in Section~\ref{s:case_study}. 

When using these data it should be noted that local solar time is not necessarily the same as the local time zone, i.e., the local midday may not be the time of day when the sun is directly overhead. Additionally in countries which use daylight savings, the difference between the local time-zone time and local solar noon will be shifted by an hour for half of the year. A solar time calculator for finding the difference between local time and solar time can be found at this link https://www.esrl.noaa.gov/gmd/grad/solcalc/. This adjustment will be made automatically for LST climatologies made via our web-tool.

Cloud cover will reduce the incoming solar radiation, and thus prevent some LST increase during the day. Enough solar radiation can pass through cloud cover to still heat the ground, so even on an overcast day there will still be a similar cycle of warmer and cooler periods of LST. However overcast days may allow extended periods of time when observations can be carried out with sufficient thermal contrast between the ground and an animal of interest.

\section{Obscuration of sources by vegetation and blending of multiple sources}\label{sec:veg_mixing}

\begin{figure}
\centering
\includegraphics[width=14cm]{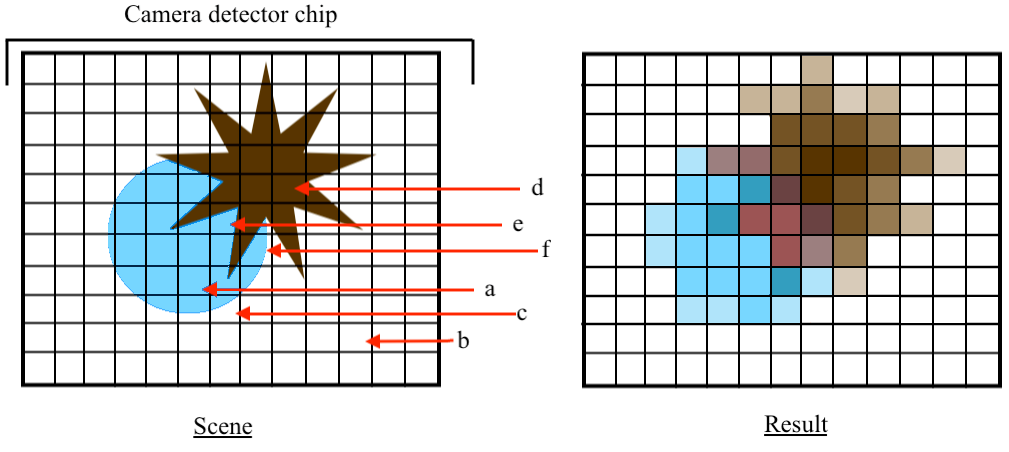}
\caption{Illustration of a field of view of a detector chip divided up into a grid to represent the pixels within the detector. The field of view contains a  target of interest (blue circle), an obscuring object (brown star), and the background (white). 
\par\noindent Top: The pixel labeled $a$ contains luminosity from only the target source. Pixel $b$ contains luminosity from only the background. Pixel $c$ contains both the target and the background, so the luminosity, and hence temperature recorded in this pixel will be a blend of that from the target source and background. Pixel $d$ contains luminosity from the obscuring object only. Pixel $e$ contains luminosity from both the target and the obscuring object, and the temperature recorded will be a blend of that from both. Pixel $f$ contains luminosity from the target, the obscuring object and the background, and the temperature recorded will be a result of the blending of the luminosity from all three.
\par\noindent Bottom: An illustration of the resulting image as would be recorded and seen when viewed in the data given the source blending.}
\label{fig:source_mix}
\end{figure}

\begin{figure}
\includegraphics[height=4.6cm]{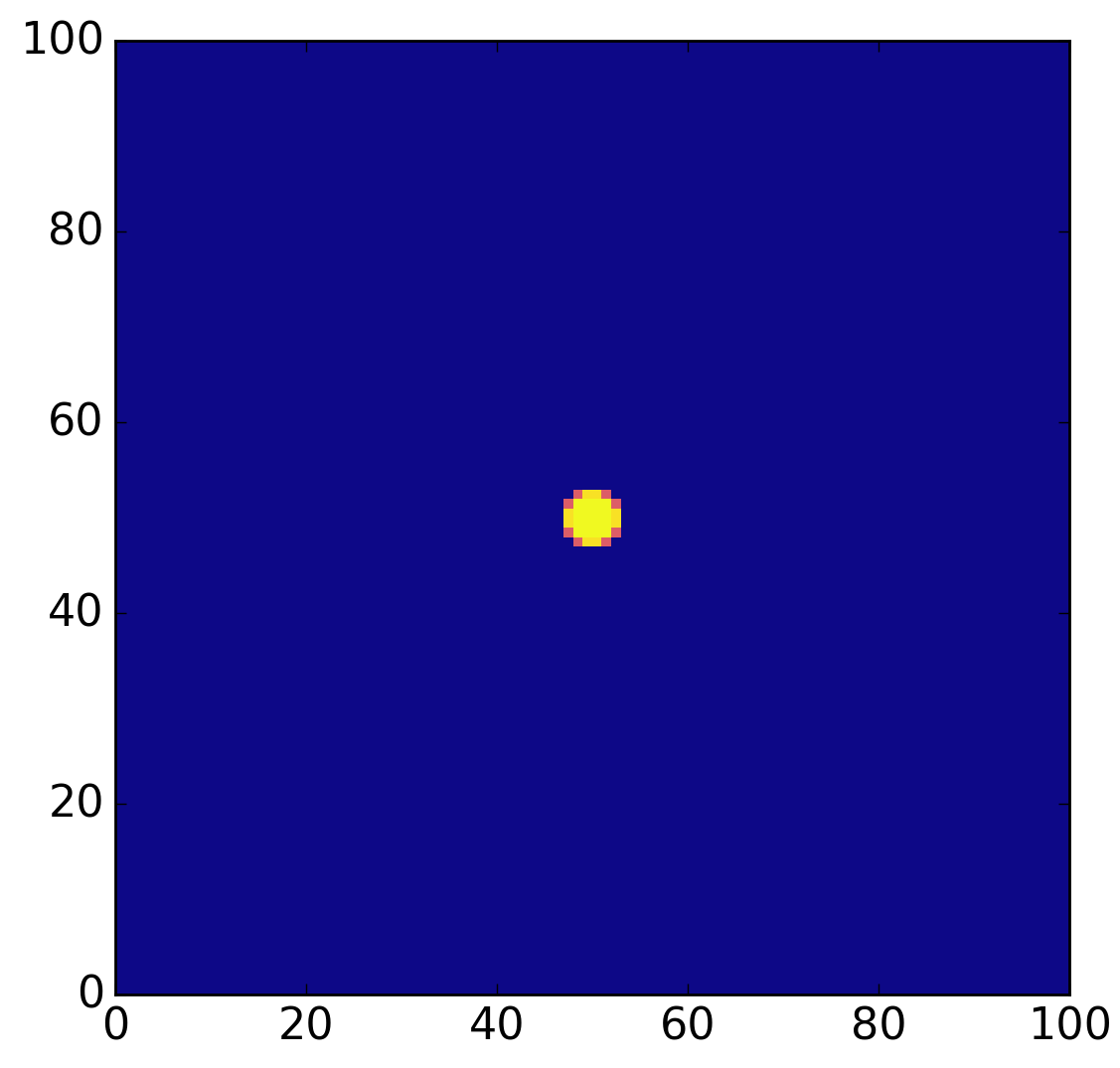}
\includegraphics[height=4.6cm]{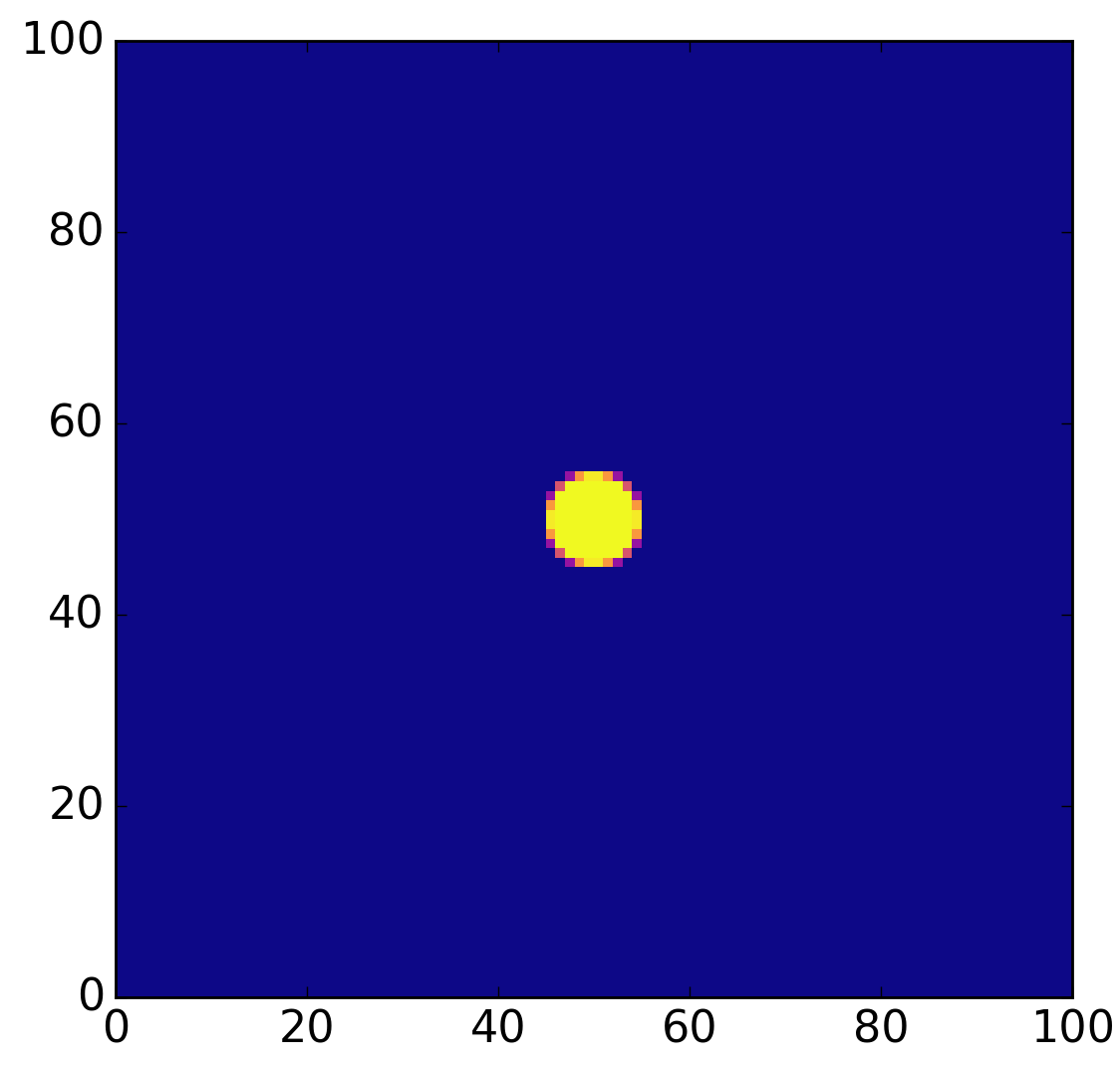}
\includegraphics[height=4.6cm]{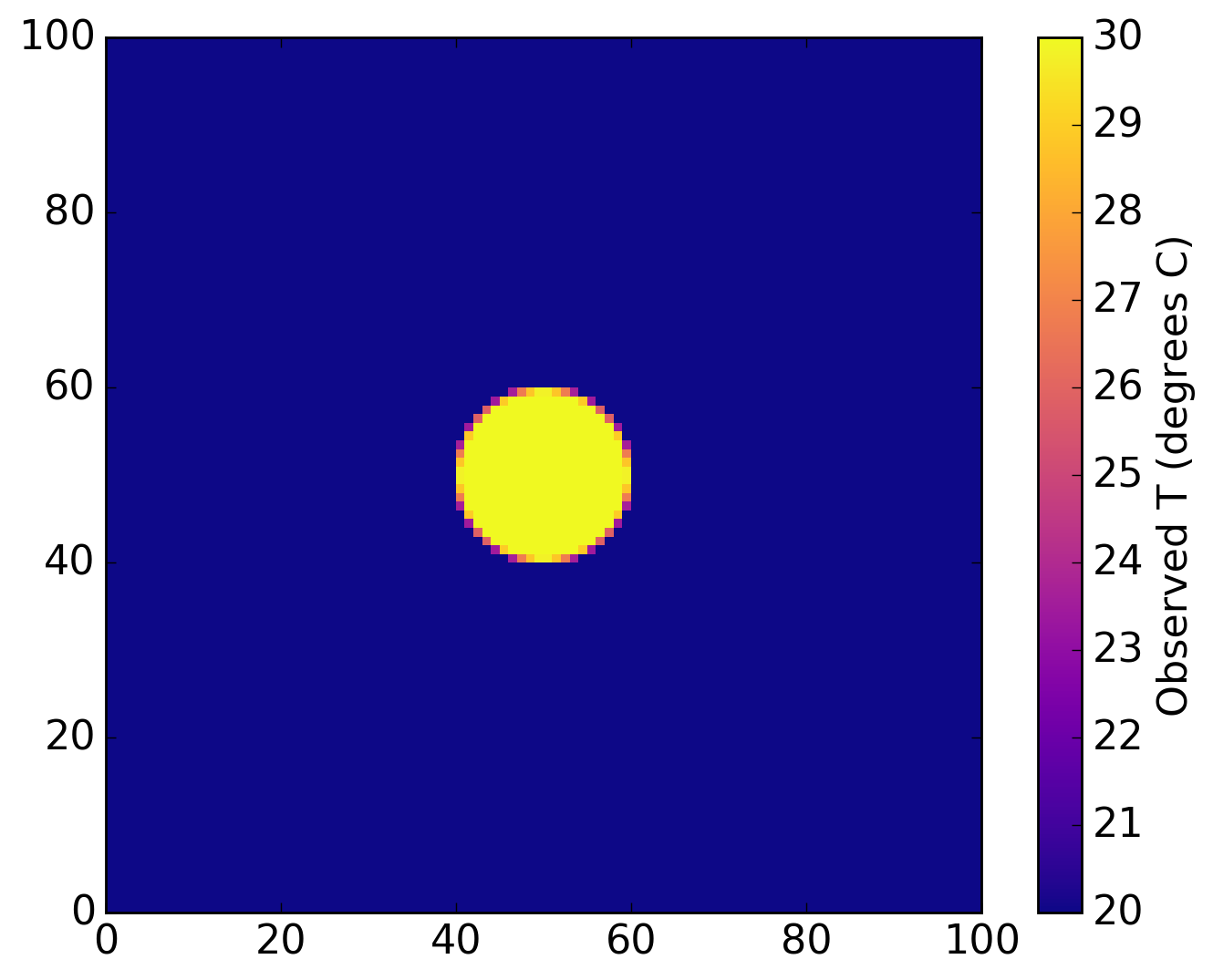}
\caption{Examples of models objects with $T_{obj}=30^{\circ}$C as would appear in a 100$\times$100 pixel detector, with diameters 3 pixels (left), 5 pixels (centre) and 10 pixels (right) in a background (surrounding region) of $T_{BG}=20^{\circ}$C. At the edges of each source the temperature recorded is somewhere between the temperature of $T_{obj}$ and $T_{BG}$ depending on what fraction of the pixel is filled by the source and background. Observed temperature per pixel calculated using Equation~\ref{equ:t_blend}}
\label{fig:spots}
\end{figure}

\begin{figure}
\centering
\includegraphics[width=8cm]{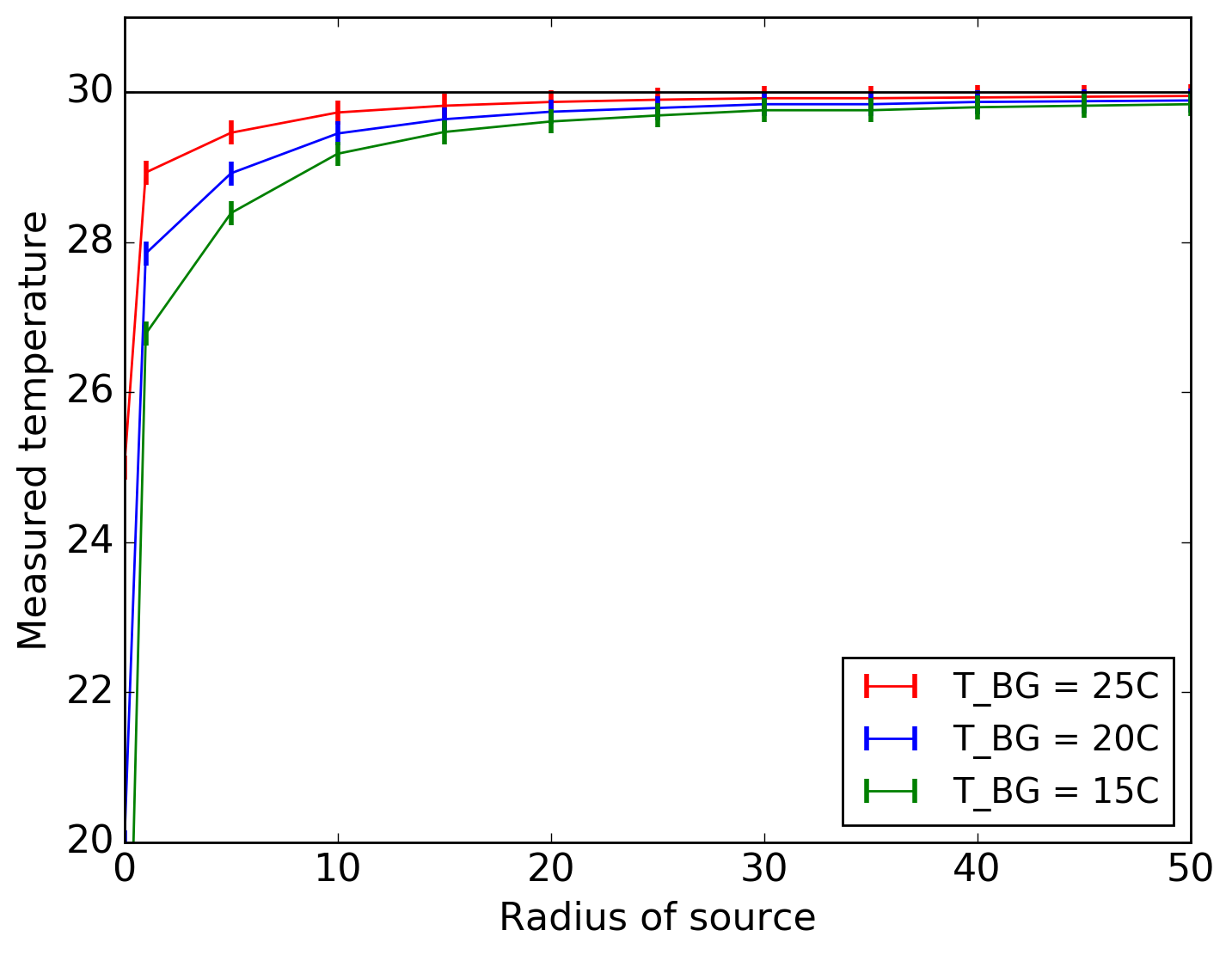}
\caption{Limitations on minimum size an object can appear in the field of view in order for its temperature to be accurately recorded. For a circular source at $30^{\circ}$C (shown in Figure~\ref{fig:spots}), the mean temperature recorded from all pixels containing the object compared with its diameter in pixels when in front of a background or within surroundings of temperature $T_{BG}$. The mean temperature recorded is within 5\% of the actual temperature when the source appears 10 pixels in diameter or larger.}
\label{spot_size}
\end{figure}

In order to detect an object in an image it must be well resolved -- i.e. large enough in the field of view to be distinguished. In addition to this, as discussed above, for thermal infrared cameras to accurately record the temperature of an object being observed the object must appear larger than a certain minimum size in the field of view. Each pixel in a TIR detector records the temperature of any objects within the view of that pixel. If more than one object is present in the space that a pixel is viewing then the temperature recorded will be a blend of the temperatures of the objects being viewed, see Figure~\ref{fig:source_mix}. For the temperature of an object to be recorded accurately in a single pixel, the light from that object must fill an entire pixel. In reality objects are seldom perfectly square and perfectly aligned with the positions of pixels in a TIR camera that might be recording them (Fig~\ref{fig:source_mix}). In order to be certain of recording an accurate temperature of an object requires there be more pixels filled by that object alone than there are pixels that are shared between the object of interest and any other sources. As the number of shared pixels decreases the accuracy of the mean temperature recorded by all the pixels increases. For a circular object projected onto a square grid of pixels the point of inflection between there being more shared pixels and more object only pixels occurs when the circle has a diameter greater than approximately 10 pixels. This is shown in Figures~\ref{fig:spots} and \ref{spot_size}.

One difficulty with observing from drones is that animals often hide beneath vegetation \citep[e.g. see][]{Mara15, Gooday18}. This can have the effect of reducing or blocking the radiation they emit. Similarly to above if the TIR emitted by an obscuring object falls within the same pixel as that emitted by the animal of interest then the temperature recorded by the detector will not be that of the animal but a blend of the animal's and the obscuring object (Figure~\ref{fig:source_mix}). 

The temperature recorded, $T_{obs}$, for a pixel containing a blend of two or more sources with temperatures $T_A$ and $T_B$ ($T_C$, ...) ({\bf temperatures in Kelvin} $\equiv$ temperature in $^{\circ}$C + 273), is described by;
\begin{equation}
T_{obs}=  \frac{T_A A_A + T_B A_B + T_C A_C + ... }{A_{total}}
\label{equ:t_blend}
\end{equation}

where $A_A, A_B$ ($A_C$, ...) and $A_{total}$ are the areas of each source A and B and the total area respectively \citep{AstroRad}.
One can calculate the observed temperature per pixel from the observed sources, background and obscuring object temperature if one knows the relative area covered by each source. 
If the temperature of the target source in the footage, when its luminosity is blended with that of an obscuring source or the background can be calculated, it is possible to determine if it would be identified as a source which is distinguishable from the vegetation or background. 

For example, using Figure~\ref{fig:source_mix}, if the target of interest (blue circle) has temperature $T_A=20^{\circ}$C, the obscuring object (brown star) has temperature $T_B=15^{\circ}$C and the background has temperature $T_C= 5^{\circ}$C. Pixels $a,~b$ and $d$ will read the temperatures of the target, background and obscuring object only, so $T_{obs}=20^{\circ}$C, $T_{obs}=5^{\circ}$C and $T_{obs}=15^{\circ}$C respectively. Pixel $c$ contains both the target and background, approximately 30\% of the pixel is covered by the object ($A_A=0.3$) and the remaining 70\% is the background ($A_C=0.7$), so the observed temperature, $T_{obs}$ recorded by the detector will be,
\begin{equation}
T_{obs}=\frac{((20+273)\times 0.3) + ((5+273)\times 0.7)}{1}.
\end{equation}
Giving $T_{obs}$=9.5$^{\circ}$C (282.5K). Pixel $e$ contains both the target of interest and the obscuring object, each source takes up approximately 50\% of the pixel. For this pixel the observed temperature will be $T_{obs}$=17.5$^{\circ}$C. Finally, pixel $f$ contains target, background and obscuring object in ratios $A_A=0.5,~A_B=0.2,~A_C=0.3$ respectively. $T_{obs}$ for pixel $f$ is given by,
\begin{equation}
T_{obs}=\frac{((20+273)\times 0.5) + ((5+273)\times 0.2)+ ((15+273)\times 0.3)}{1},
\end{equation}
$T_{obs}=15.5^{\circ}$C (288.5K). In most real world cases it will likely be impossible to tell the exact ratios of different sources within each pixel. The ratios of target : background or target : obscuring vegetation will vary depending on the shape and size of the target and the type of vegetation (or other object) which is obscuring it. In the case of vegetation, sparse vegetation may only cover a small fraction of the target of interest, whereas dense of very leafy vegetation may obscure a large fraction of the object. As discussed in Section~\ref{sec:fov}, the size of the target will dictate the drone flying height. For accurate data the observing strategy will also need to be adjusted to account for the variety of vegetation which may be obscuring the targets, possibly flying lower if the fraction of the target which may be covered by vegetation is high.
The differences in temperatures recorded by the blending of multiple sources within the same pixel highlights the need for targets of interest to appear at least 10 pixels in diameter in the field of view when taking data if one wishes to record an accurate temperature for the target and distinguish it from the surrounding background and other sources.

\section{Case study: riverine rabbit}\label{s:case_study}
We now show how the above understanding of the atmosphere, camera field of view and climatology can be used to improve the efficiency of TIR monitoring of animals. In particular we applied this to a data-collecting field trip to try to detect the riverine rabbit ({\it Bunolagus monticularis}). Residing in the vast arid area in the central part of South Africa commonly referred to as the Karoo, the riverine rabbit is one of the most endangered, and rarely observed mammals on the planet (see https://www.iucnredlist.org/species/3326/43710964). The rabbit is found across a region the size of Austria, approximately 54,227 km$^2$, and it is estimated to have a very small population of approximately 200 adult individuals \citep[see][]{CollinsToit}. -- however this number is very difficult to estimate accurately due to the species' low density distribution. The riverine rabbit's main habitat is near rivers where it feeds on and shelters beneath the riparian vegetation. Being critically endangered, spread over a wide area, and notoriously shy, it is seldom spotted in foot surveys which are the traditional method of monitoring populations (however some progress has been made with camera trap surveys -- C. Theron, private communication). As such, this species represents a prime target for monitoring with drones.

The drone system used was a DJI Matrice 600 with a FLIR T640 TIR camera mounted on it. The camera we used has a 13mm lens, with FOV 45$^\circ\times$37$^\circ$, detector size 640$\times$512 pixels, and absolute temperature sensitivity $\sim\pm$5\%.

To implement the methods described above in order to observe riverine rabbits, we followed the steps below;
\begin{itemize}
\item For the desired latitude and longitude calculate the LST annual climatology and select the best time of year to fly -- ideally the coolest time of year for best thermal contrast. We also recommend checking the typical annual rainfall cycle for the year to avoid rainy seasons (this is included in the web-tool).
\item For the time of year selected examine the LST daily climatology and select the coolest time of day for flying.
\item Given the size of the target animal to be observed, calculate the optimum height to fly the drone for the necessary data quality.
\item Calculate the atmospheric effect for the predicted weather conditions. This is especially worth checking if flying at heights greater than 100m AGL or with the camera at an angle.
\item Estimate vegetation coverage for how much targets are likely to be obscured, and whether they will be able to be distinguished from the ground. If animals of interest are known to hide beneath vegetation then adapt observing strategy accordingly.
\item Consider the angle at which the camera should be mounted -- straight down will allow for simpler analysis, but angled setups may allow some effects of vegetation to be circumvented.
\item An easy method for calculating all these steps can be found in our web-tool -- \url{http://www.astro.ljmu.ac.uk/~aricburk/uav_calc}.
\end{itemize}

Riverine rabbits are often found near Loxton, South Africa -- latitude: $-31.476421$, longitude: $22.354109$. 
We calculated the annual LST for Loxton (see Figure~\ref{fig:loxton_lst}) and found that the coolest time of year was between May and September. 
We collected data between 21st--27th September 2017, and used the LST calculator to produce a climatology for the time of year -- shown in the right panel of Figure~\ref{fig:loxton_lst}. We assumed the external temperature of the riverine rabbit to be similar to that of a domestic rabbit. To estimate this we observed a domestic rabbit using the same camera and drone setup at a range of heights between 15--30 meters, and found it to be between $T_{rabbit}=20-30^{\circ}$C. 
From Figure~\ref{fig:loxton_lst}, the ground temperature will typically exceed 20$^{\circ}$C between 0800 and 1900 -- meaning the best time for observing was likely before 0800. The observed ground temperature between 0600 and 0800 when we carried out our observations was $4-13^{\circ}$C.

We took data from flights at several times of day -- 0600, 0700, 0800, 0900, 1800 and compared these with the climatology. For a rough but simple to implement estimate, we assumed that the 50\% of coolest pixels in the footage from each flight would be background, and that any sources which correspond to animals, humans (etc) would be in the warmest 50\% of pixels. We took all the average temperature value for all the pixels in the bottom 50\% in all the frames from each flight as the average LST, and two standard deviations ($2\sigma$) from the mean to be the uncertainty on the LST.
Figure~\ref{fig:loxton_lst} shows the comparison between the observed and climatological LST values. During September at the latitude and longitude of Loxton, the local solar time is local time zone $-37$ minutes. In the figure the recorded times are adjusted accordingly.  As would be expected from the climatology, the ground heats up very quickly as the daytime progresses and after 0800 it becomes very difficult to distinguish animal sources from the ground, see for example Figure~\ref{fig:lst_comparison} at 0800 and 0900. 

The riverine rabbit is estimated to be between 25--30cm in length. Given Equation~\ref{equ:pix_scale}, and the resolution of our FLIR Tau 640 camera, we calculated the optimum height to fly in order for the rabbit to appear at least 10 pixels in the field of view of the camera and to cover the largest area of ground whilst observing, was $h=20-24$ meters.

Given the location of Loxton on a high escarpment, and the cool, dry nature of the region early in the morning, atmospheric conditions were generally good for observing with minimal atmospheric effects on the infrared emission from potential targets. We measured the atmospheric conditions with a Kestrel Weather Meter, and used this to confirm the LST recorded by the TIR camera. Throughout the course of the observing time, the air temperature, pressure and humidity were 5--20$^{\circ}$C, 95--100kPa and 25--70\% respectively. At a drone height of 20 meters this meant that only $\sim$0.5--1\% of the thermal infrared from sources was absorbed. Using Equation~\ref{equ:dT}, the temperatures recorded for riverine rabbits would be $T_{obs}=19-29^{\circ}$C. With the ground temperatures recorded above ($4-13^{\circ}$C), we would expect the observed temperature of the ground to be $4-14^{\circ}$C, meaning that riverine rabbits should be distinguishable from the surroundings.

\begin{figure}
\centering
\includegraphics[width=9cm]{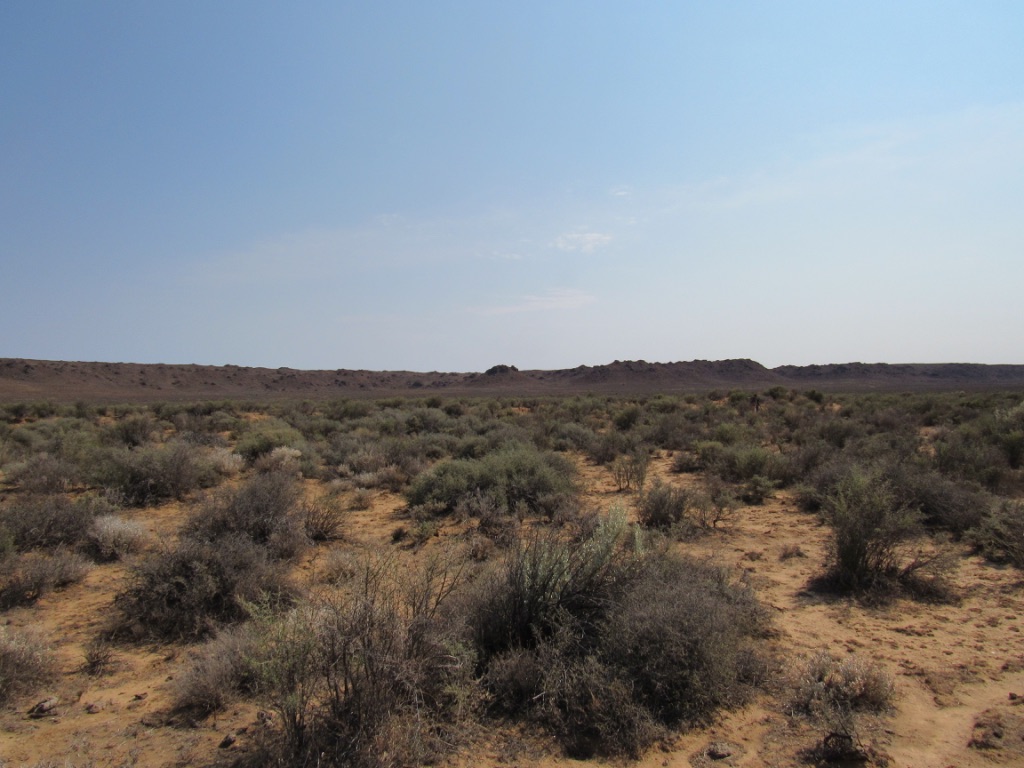}
\caption{An example of the terrain and vegetation in the Karoo region. The height of the vegetation depicted is between $\sim$0.5--1.0 meters, and the widths or individual shrubs are in the range 0.3--1.0 meters. The spacing between shrubs is irregular and between 0.5--5.0 meters.}
\label{fig:karoo_pic}
\end{figure}

The vegetation in the Karoo region is largely made up of small, dense shrubs. From ground level it is not possible to see further than a meter or two through the bush. This further motivates the use of a drone for surveying, as it is easy to miss sightings of small animals when on foot, even if one passes close by the animal. Individual bushes, however, are also dense, obscuring much of the emission from any animal hiding beneath them, see Figure~\ref{fig:karoo_pic} for a typical example of the terrain as viewed from ground level. Assuming the external temperature of a rabbit to be 20$^{\circ}$C, and the temperature of the vegetation was measured to be 5$^{\circ}$C (early morning), and with the bush being fairly dense so obscuring 75\% of the surface area of anything beneath it, using Equation~\ref{equ:t_blend} the measured temperature of a rabbit beneath the bush would be 8.75$^{\circ}$C. Whilst the difference between the observed temperature of rabbits should be distinguishable from the vegetation for a camera with relative thermal sensitivity of 0.05$^{\circ}$C, the effect of the vegetation is to cause the target to have a much smaller difference in temperature from its surrounding than might be initially expected. This smaller difference in temperature must be considered when it comes to visual inspection of the data for detection of riverine rabbits, as by eye it may be easy to miss targets of interest as a result.

\begin{figure}
\includegraphics[height=6cm]{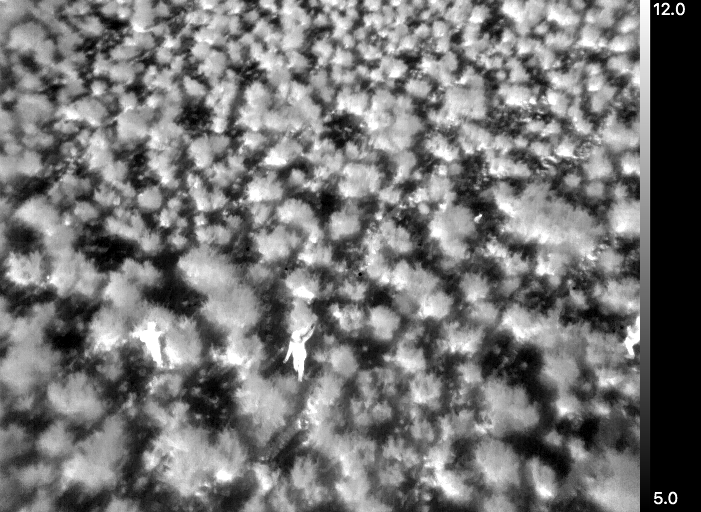}
\includegraphics[height=6cm]{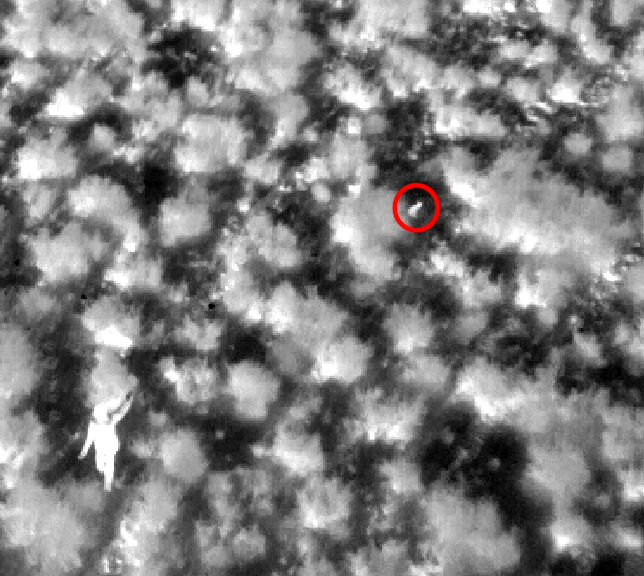}
\caption{Example of frame containing riverine rabbit. Sighting occurred at 0715 on 25/09/2017. The ground temperature was LST $\simeq 5^{\circ}$C and the air temperature was $\simeq 9^{\circ}$C. Left shows entire field of view, right shows a zoom in on the riverine rabbit, with the rabbit indicated by the red circle.}
\label{fig:rabbit}
\end{figure}

Since the riverine rabbit is known to rest under the edges of the bush \citep{CollinsToit}, we decided to mount the camera with a 45$^{\circ}$ inclination to allow us to look under the bush and have a better chance of sighting them. For the same resolution as  $h$=20 meters straight down (10 pixel resolution for rabbits), for the centre of the field of view we need to have $R_M$=20 meters at an angle, and the flying height calculated from Equation~\ref{equ:any_R} is $h=14$ meters. Since the minimum safe flying height was 15 meters, we flew at 15 meters. The variation in pixel size with distance along field of view ($D_F-D_C$ in Figure~\ref{fov_fig}, see Appendix~\ref{app:angle_math}) meant that the bottom of the field of view was $W_C=13.9$ meters wide along the $x$-axis, and the top was $W_F=28.3$ meters wide. Since the camera is at an angle, at the top of the y-axis of the field of view it will be mostly filled with vegetation, so whilst this setup allows us to see beneath the vegetation in the middle section of the field of view, large parts of the top of the field of view will still be obscured by vegetation (e.g., see Figure~\ref{fig:rabbit}). 

The low flying height necessary for resolving the relatively small riverine rabbit is potentially a major limitation when considering surveying large areas. Flights at 15 meters AGL, giving a relatively small FOV on the ground, would be very limiting for covering large areas in the typical flying time afforded by multirotor-style drone batteries. 15 meters AGL is also too low for large area surveys by fixed wing drones, which typically require higher flying heights for safety reasons. One solution to this would be to use a camera with a longer lens, giving a finer angular resolution and thus allow a drone to be flown higher whilst still resolving riverine rabbits. When observing larger species than rabbits, prohibitively low flying heights will be less of an issue.

There were few spurious sources in the region, as early in the morning most other warm objects are likely to be other animals, which are not the target of interest. We made one surprising discovery, that the burrows made by animals (e.g. aardvarks, bush rats, gerbils) in the area retained heat overnight and could occasionally appear as warm sources early in the morning before the surface ground had had time to heat up. Fortunately the shapes of the burrows were much different to animals, making them easy to account for. 
Another affect worth noting is the differential heating of objects as one side is illuminated at sunrise and sunset, making bushes (etc) appear warmer on one side, and casting a thermal shadow on the other in the TIR footage.

We carried out the initial flights with some trained human observers walking below the drone to confirm any sightings. We observed with the drone for 7 days, making 2 flights each morning, and recorded 5 sightings of riverine rabbits, with footage of two. Figure~\ref{fig:rabbit} shows a frame from one of the sightings, a video of this sighting can also be found at \url{http://www.astro.ljmu.ac.uk/~aricburk/riverine_rabbit.html}.

\section{Broader applications of methods}\label{lit_comparson}

Whilst the use of thermal-equipped drones is still in its early and experimental stages as a tool for ecological studies and conservation, it is important to reflect on advances made so far and how the application of new methods may benefit future studies. To illustrate the benefits of using the methods we have presented we examine some examples of previous work in the literature, and suggest how the methods we present may be applied in planning similar studies in future.

\cite{Christiansen14} describe an automated detection and classification system using TIR for small animals hiding in agricultural fields. \cite{Isreal11} also describe a prototype for a similar system for detecting small animals in agricultural fields. The kinds of challenges these groups faced to the success of their systems were uncertainties in the best height to position the camera for a good trade off between area coverage and resolving small animals, and blending of thermal radiation from sources of interest and the ground or vegetation. 
The calculation of the pixel scale of the camera used, and how this changes with  height will allow studies like these to know in advance the maximum height to fly for a good trade off between area covered and minimum observed size of targets of interest. This would help ensure the reliability of a detection system of this type, and would aid in classification and tracking of targets found. Understanding expected source temperatures as a result of blending with vegetation would also be beneficial for accurate detection and identification. Knowing in advance the daily ground climatology would also allow for effective planning of the time of day to carry out a similar survey.

In a similar study \cite{Israel17} test the effectiveness of TIR and RGB cameras mounted together on a drone to find lapwing nests in agricultural fields so that they can be avoided when harvesting or mowing. They rely on the heat given off by the eggs in the nests to spot them in the TIR footage and see them well below a certain height. They do not see the eggs at all in their RGB footage, demonstrating the major advantage of thermal imaging even for objects as small as lapwing eggs ($\lesssim 5$cm diameter). Since the eggs are small, a very low flying height would be needed for 10 pixel resolution, around 4 meters. However, since eggs are simple, being almost round and uniformly warm across their surface, and usually much warmer than their surroundings due to their lack of insulation, if knowledge of source blending was applied one could estimate how the eggs would appear even if not well resolved, especially if more than one egg was present in the nest (as \cite{Israel17} report is indeed the case).

\cite{Chretien15, Chretien16} test a range of automatic systems for detecting animals in captivity (deer, elk, bison and wolves) with duel RGB/TIR camera system on a drone, like \cite{Israel17} they find that thermal object detection largely more accurate than RGB. However they also find confusion with spurious warm sources (rocks etc) and the heat of the ground during the day (flights performed between 0700 and 1300 local time) making it difficult to distinguish animals from the ground.
\cite{Witczuk17} survey populations of ungulates in forests using TIR. In this study they determine the best heights and time of day for flying the drone to get high signal-to-noise detections of the ungulates through trial and error. The use of trial and error for finding optimal flying heights and times of day is common \citep[e.g.][]{Mara14} and still has a place in optimizing observing strategies, even when applying the pre-planning methods we describe above. Inevitably, in the field, situations will be more complex in reality than these simplified methods and calculations can account for, however the results generated by the application of these methods can provide a good stating point and save a lot of time which might be spent with trial and error observations, allowing for more efficient use of valuable field observing time and subsequent data analysis time.
For optimum efficiency for studies similar to these, we recommend using the methods described here for height and physical pixel scale calculation, and the LST climatology for the region they survey. The physical pixel scale of the data for given heights would also make different species easier to identify by knowing what physical size each detected animal corresponds to.

\cite{Lhoest15} use TIR and an algorithm to count hippos within groups. The algorithm uses a thresholding method -- whereby hippos are detected by having a temperature above a certain level compared to their surroundings. This kind of study has great potential in advancing the automation of detecting animals, monitoring their numbers, and gaining very thorough knowledge of the status of wild populations of animals.  
For this kind of study knowing the physical pixel scale in advance is vital for being able to optimise the drone flying height and knowing how large hippos would be expected to appear in the data. Additional information about the expected temperature of the hippos in the data as a result of blending of temperatures within pixels containing both hippos and water at the edges of the hippos, would also allow the thresholds to be set empirically rather than by trial and error, potentially allowing for very accurate animal counts.

\cite{Burn09} use TIR in an aerial survey by plane to count walrus groups. The advantage of such a survey is that a very large area can be surveyed - hundreds or thousands of square kilometers, allowing sparse populations in isolated areas to be monitored. 
At the height of manned aircraft, absorption by the atmosphere will have a large affect on the luminosity of TIR radiation received. By knowing the expected temperature change between emitted and received radiation as a result of atmospheric absorption future studies can have reliable knowledge about what to look for when detecting animals. As with studies similar to \cite{Lhoest15}, high detection accuracy will benefit from knowledge about the expected temperature of the groups where animal and background source blending is occurring (at the edges of groups or in the gaps between group members).

\section{Conclusions}
We have outlined methods to optimize a strategy for observing homeothermic animals using thermal infrared cameras mounted on drones, and account for some of the environmental issues associated with thermal infrared imaging. We have shown using a case study that these methods are informative and can be applied to make best use of available time when observing in the field and allows some trial and error optimization to be avoided.

We tested each part of our method with a case study to try and detect the very rare and endangered riverine rabbit in Loxton, SA. We found that the observed land surface temperature (LST) was similar to that predicted by our LST climatology. Since riverine rabbits can be seen in the footage obtained, we conclude that the flying height calculated was sufficient for resolving riverine rabbits. However since all sightings were confirmed by an observer on foot, we are not yet able to assert that riverine rabbits could be distinguished from other species of similar size and shape (e.g.Scrub hare -- {\it Lepis saxatilis}, and red rock rabbit -- {\it Pronolagus spp.}, are also found in the area) purely using TIR data. The atmospheric absorption for the environment in our case study was small, and after correcting for atmospheric effects we recorded the LSTs with the thermal infrared camera to within $\pm 0.2^{\circ}$C of a thermometer on the ground. We calculated that the rabbits would be difficult to distinguish from the vegetation when viewed from above so we adopted a strategy of flying with the camera at an angle. The area we were searching was known to be home to a small population of rabbits, and following the strategy devised from 14 separate flights we observed a total of 5 rabbits. These sighting are the first thermal infrared data of riverine rabbits recorded in the wild.

We have also discussed how the methods we have presented could be used to build on the work of past studies, and we hope that other groups will find these methods useful in planning their observations. The methods we present here will be made available to the community through an easy to use web-tool (\url{http://www.astro.ljmu.ac.uk/~aricburk/uav_calc/}).

\section{Acknowledgements}
We thank staff of the Drylands Conservation Programme of the Endangered Wildlife Trust, Lourens Leewner, Bonnie Schumann, Matt Pretorius, Este Matthew and Dr. Ian Little for their assistance with data gathering and providing invaluable expertise on the riverine rabbit.
We thank Mara Mulero-Pazmany for her advice and input on the style and content of this paper.
We acknowledge Lizzie Good, Rob King, Colin Morice at the UK Met Office for advice on using land surface temperature data and MODIS data. 
We thank Andy Goodwin for producing drone mounts and providing technical expertise.
This research was funded by a ISTAT Foundation Humanitarian Grant. C.B. is funded by STFC Global Challenges grant.

%
%
%
%
%
%
%
%
%
%
%
%
%
%


\bibliographystyle{tfcad}
\bibliography{references}

%
%
\appendix

\section{Field of view and drone height with an angled camera}\label{app:angle_math}

In Section~\ref{sec:fov} we describe the relation between distance of the camera from the target of interest and its recorded size in pixels in the resulting data that may be recorded. For a detector mounted on a drone at nadir position, the angular pixel scale, $\rho_{a}$ is given by,
\begin{equation}
\frac{\theta}{\#pixels}= \rho_{a},
\label{app_equ:ang_pix_scale}
\end{equation}
where, $\theta$ is the field of view (FOV) in degrees, and $\#pixels$ is the number of pixels in the detector. The resulting physical pixel scale,  $\rho_{p}$ as a function of drone height, $h$, is given by; 
\begin{equation}
h~tan(\rho_{a})= \rho_{p}.
\label{app_equ:pix_scale}
\end{equation}

When flying a drone with a camera at an angle the FOV will be distorted from the rectangular shape seen when the camera is pointed straight down. In this case the distance between the drone and different points in the camera FOV will vary, as will the projected size of each pixel on the ground.
Following the notation in Figure~\ref{fov_fig}, with $\theta_x$ and $\theta_y$ being the angular field of view of the camera in degrees in the $x$ and $y$ directions respectively, and $\phi$ being the angle between the centre of the camera FOV in the y-axis and the nadir, the size of the field of view on the ground in the $y$-direction (vertical from the point of view of the camera) is given by;
\begin{equation}
D_C = h.tan\left(\phi - \frac{\theta_y}{2}\right),
\end{equation}
\begin{equation}
D_F=h.tan\left(\phi+\frac{\theta_y}{2}\right),
\end{equation}
\begin{equation}
D_F-D_C = h\left(tan\left(\phi+\frac{\theta_y}{2}\right)- tan\left(\phi-\frac{\theta_y}{2}\right)\right).
\label{equ:fov_angle_y}
\end{equation}
The projected FOV in the $x$-direction (horizontal as viewed by the camera) varies as a function of distance along the $y$-direction, so that $W_C,~W_M$ and $W_F$ are given by;
\begin{equation}
W_C=2(D_C^2+ h^2)^{1/2} ~tan\left(\frac{\theta_x}{2}\right),
\end{equation}
\begin{equation}
W_M=2(\left(\frac{D_C+D_F}{2}\right)^2+ h^2)^{1/2} ~tan\left(\frac{\theta_x}{2}\right),
\end{equation}
\begin{equation}
W_F=2(D_F^2+ h^2)^{1/2} ~tan\left(\frac{\theta_x}{2}\right),
\end{equation}
where $\theta_x$ is the angular FOV of the camera in the $x$ direction. Similarly $\theta_y$ is the angular FOV in the $y$ direction.
The values of $R_C,~R_M,~R_F$, for each of $D_C,~D_M,~D_F$, are given by,
\begin{equation}
R=\sqrt[]{(h^2 + D^2)}.
\label{equ:r_vs_d}
\end{equation}
The physical pixel scale for this set up is thus variable as a function of location in $x$ and $y$ within the field of view, and is given by;
\begin{equation}
\rho_p= R.tan(\rho_a), 
\label{equ:pix_scale_angle}
\end{equation}
for each of $R_C,~R_M,~R_F$ and for $\theta_x$ and $\theta_y$. For calculating the physical pixel scale for each specific pixel in the field of view, we must calculate the value of $R$ for the $N$th pixel along the $y$-axis, where $N=1$ is the bottom of the field of view. This relation is described by,
\begin{equation}
h=R.Cos\left(\left(\phi-\frac{\theta_y}{2}\right)+\rho_aN\right).
\label{equ:any_R}
\end{equation}
Thus the physical pixel scale for any pixel in the field of view can be calculated using Equation~\ref{equ:pix_scale_angle}. The equation can also be used to calculate the flying height in order for a target to appear the desired size within the middle of the field of view. 

In the example discussed in Section~\ref{sec:fov}, $\phi=60^{\circ}$  and $\theta_y =37^{\circ}$, in this case the top of the field of view is at $\phi + \theta_y/2$=78.5$^{\circ}$. In cases where $\phi + \theta_y/2>90^{\circ}$ the top of the field of view will be above the angle of the horizon (90$^{\circ}$). For fields of view which extend above the horizon, clearly the above calculations do not hold.

It should be noted that for large values of $h$ with the camera mounted to point straight down, the distance to the edges of the field of view ($R$) will be larger than $h$, meaning the pixel scale will not be uniform across the entire field of view. See Figure~\ref{corrections_arrays} for an illustration of this.

\section{Radiative transfer of TIR radiation from a source through the atmosphere}\label{app:rad_tran}

In Section~\ref{sec:absorb} we discuss how the properties of the atmosphere through which thermal infrared radiation must pass before reaching the detector will affect the temperature that is recorded, a typical demonstration of this is illustrated in Figure~\ref{abs_coeff}. The change in intensity of radiation as it passes through a given medium is given by the equation of radiative transfer;
\begin{equation}
\frac{dI_\nu}{dS} = -\kappa_\nu I_\nu + j_\nu,
\label{equ:rad_tran}
\end{equation}


\noindent
where $I$ is the intensity of radiation at frequency $\nu$ from the source, $S$ is the distance that the light has traveled from the source through the intervening medium, $\kappa$ is the absorption co-efficient of the medium, and $j$ is the emissivity of the medium. For conditions typical of the lower atmosphere ($<1$km above sea level) at typical drone flying heights (up to a few hundred meters), $j_\nu$ can be approximated by, 
\begin{equation}
j_\nu = \kappa_\nu B_\nu(T),
\label{equ:rad_tran_int}
\end{equation}
where $B_{\nu}$ is the Planck function.
%
%
Equation~\ref{equ:rad_tran} then integrates to,
\begin{equation}
I_\nu (S) =I_\nu (S_0) e^{-\kappa_\nu S} + B_\nu(T)(1-e^{-\kappa_\nu S} ),
\label{equ:rad_tran_int}
\end{equation}
where $I_\nu(S_0)$ is the intensity of the source as distance $S=0$ and $I_\nu(S)$ is the intensity of the radiation from the source at distance $S$ (see Figure~\ref{abs_coeff}). For calculating the fraction of radiation from a source which is received by a detector, we rearrange for $I(S)/I(S_0)$,
\begin{equation}
\frac{I_\nu (S)}{I_\nu (S_0)} = e^{-\kappa_\nu S} + \left(\frac{e^{h\nu/kT_{s}}-1}{e^{h\nu/kT_{m}}-1}\right)(1-e^{-\kappa_\nu S} ),
\label{equ:frac}
\end{equation}
where $h, ~k$ are Plank's constant and Boltzman's constant respectively, and $T_s, ~T_m$ are the temperature of the source and medium respectively. This can be converted to an observed temperature, $T_{obs}$ by,
\begin{equation}
T_{obs}=T_{s}\left(\frac{I_\nu (S)}{I_\nu (S_0)}\right)^{1/4}.
\label{equ:dT}
\end{equation}
 See \cite{DysonWilliams} for full details of the equation of radiative transfer and its application.

\subsection{Application to observing with TIR cameras on drones}\label{app:abs_spec_sec}

\begin{figure}
\centering
\includegraphics[width=9cm]{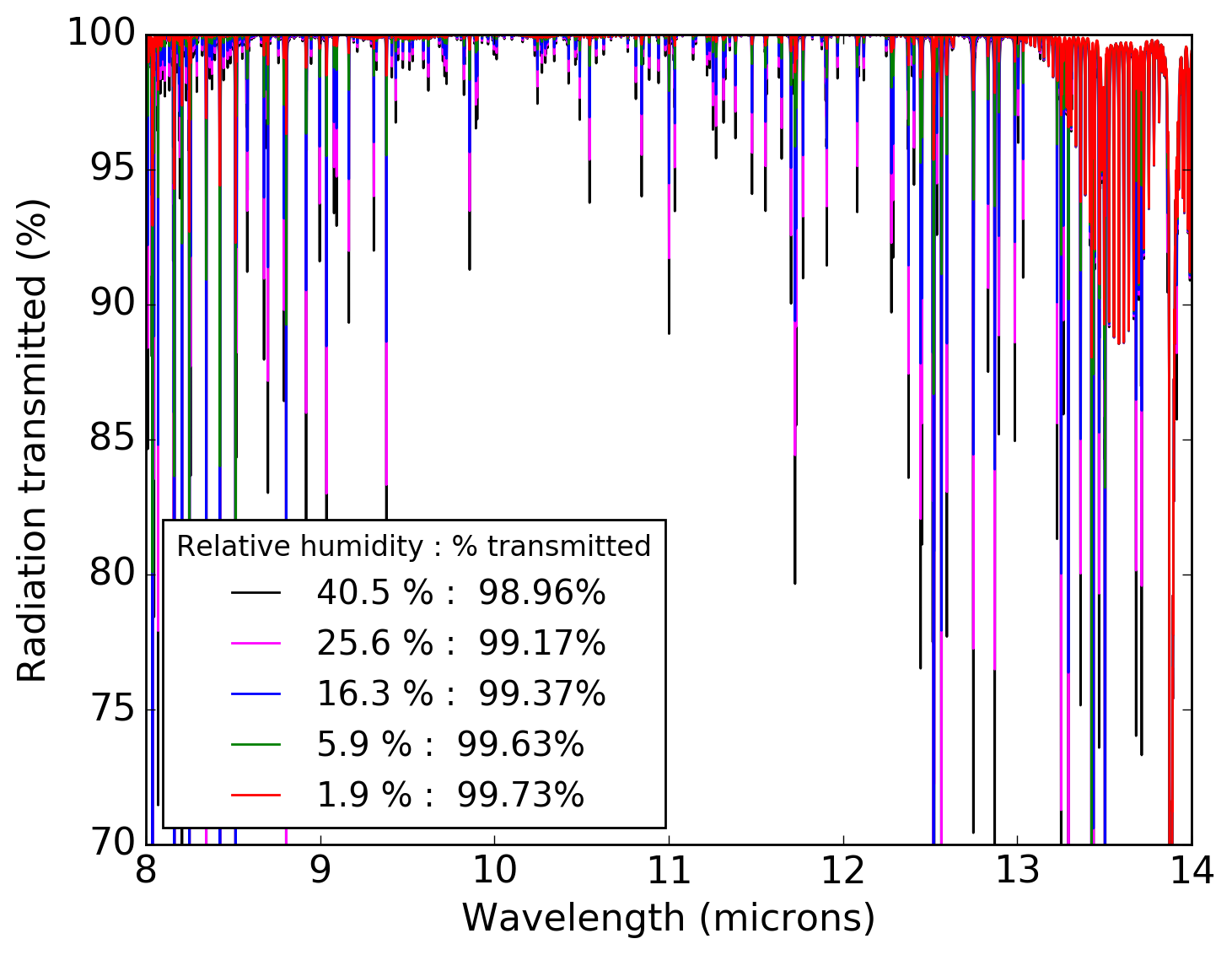}
\includegraphics[width=7cm]{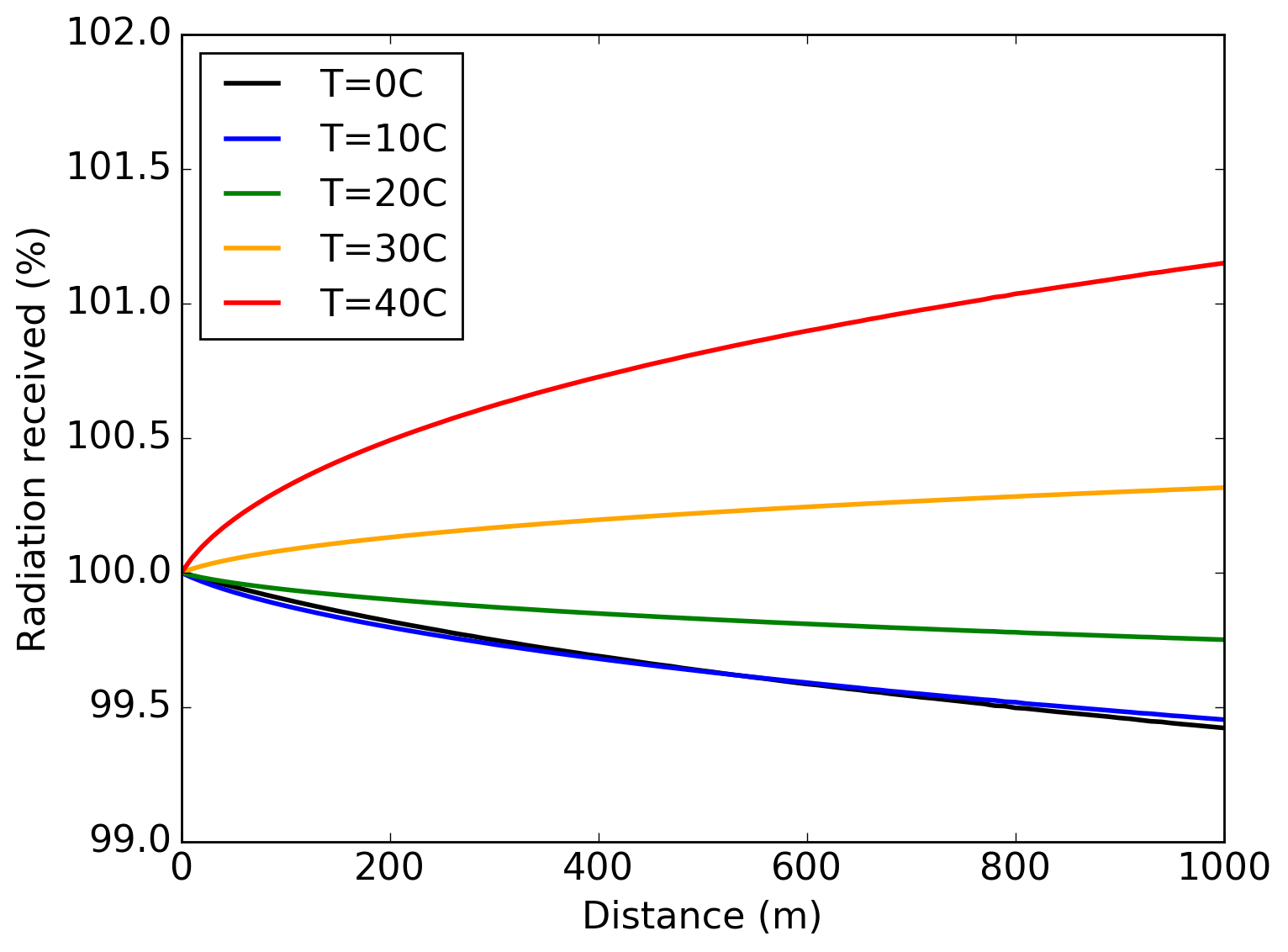}
\includegraphics[width=7cm]{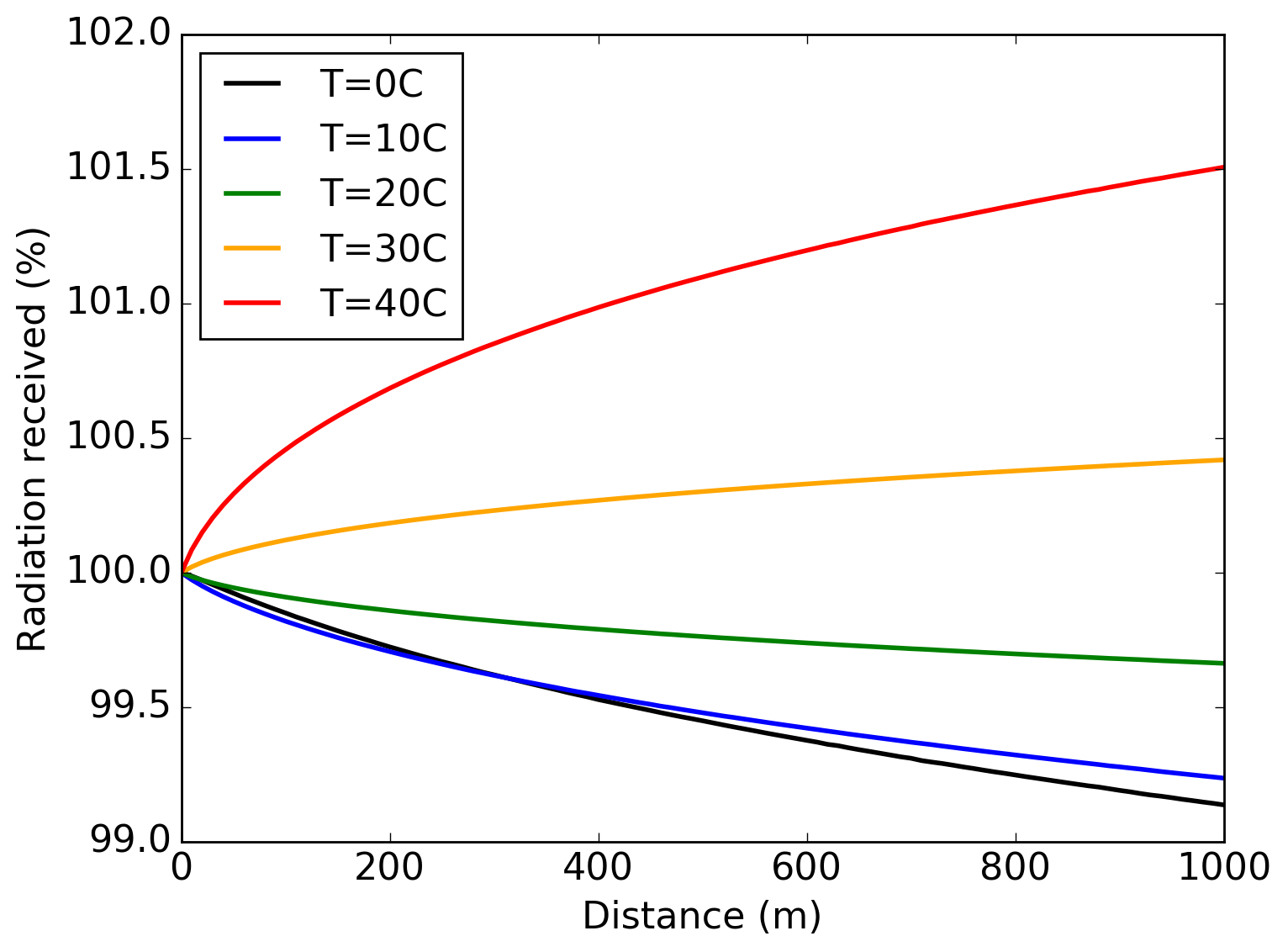}
\caption{Top: Absorption spectra for atmospheric gases at 1 atmosphere pressure (101.325 kPa) and temperature of $T_m=40^{\circ}$C (313K) for different values of relative humidity. The integrated fraction of emitted TIR which is transmitted over the whole spectral range is indicted in the legend. 
Bottom left: Integrated percentage of flux from source transmitted for different air temperatures ($T_m=0-40^{\circ}$C $\equiv$ 273--313 K) vs distance from source for atmosphere at $\sim$1 atmosphere pressure (101 kPa) and 50\% humidity. 
Bottom right: The absorption and emission by atmospheric gases in a more extreme atmospheric scenario (similar to that which might be found on a hot day in a tropical rain forest) of $T_m=40^{\circ}$C air temperature, 110 kPa pressure and 100\% humidity.}
\label{app_fig:abs_spec}
\end{figure}

To determine the potential effect of atmospheric absorption on the TIR measurement made with thermal cameras on drones we take data from the HITRAN molecular spectroscopic database \citep[][for full details]{hitran, hitran_b} for a mixture of gases representative of Earth's atmosphere at various values of humidity, pressure and temperature. From these absorption spectra can be generated for different combinations of atmospheric temperature, pressure and humidity; an example is shown in Figure~\ref{app_fig:abs_spec}. By integrating under the curve it is possible to calculate the resulting change in intensity of flux from an object with distance. The changes in intensity are converted into a change in temperature using Equation~\ref{equ:dT}, and shown in Figure~\ref{fig:spectr_abosorb}.

\begin{figure}
\includegraphics[width=7.5cm]{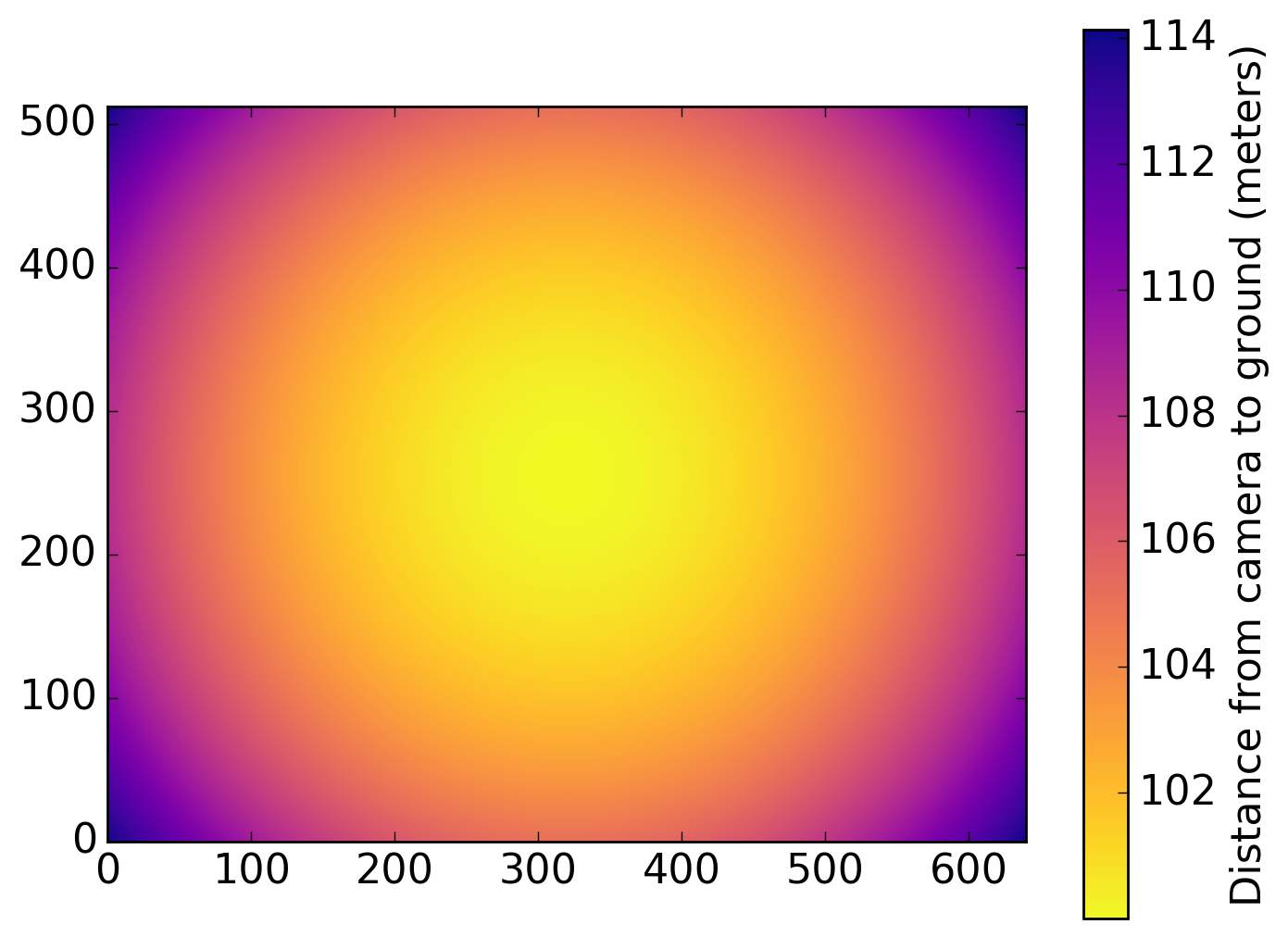}
\includegraphics[width=7.5cm]{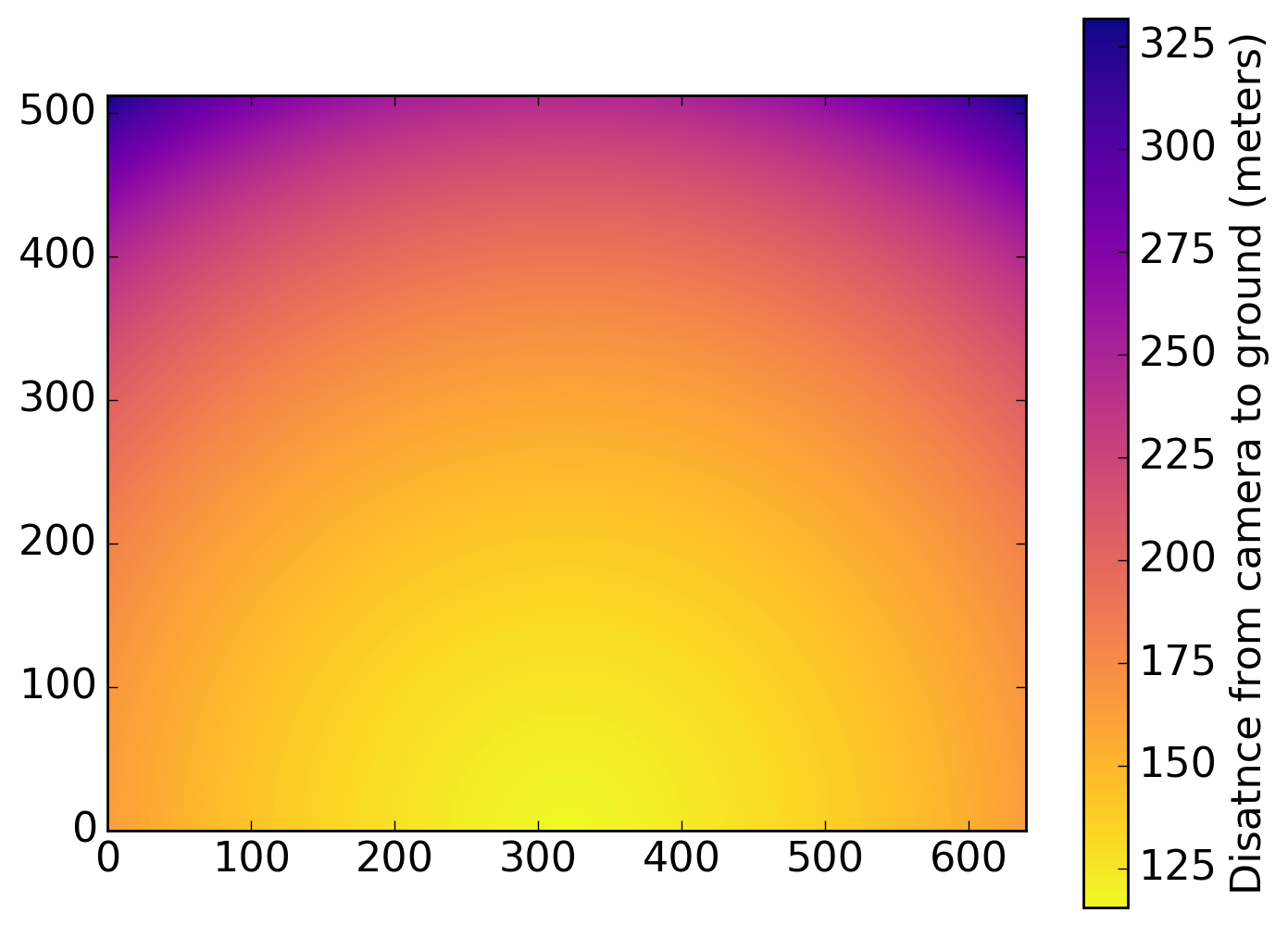}
\includegraphics[width=7.5cm]{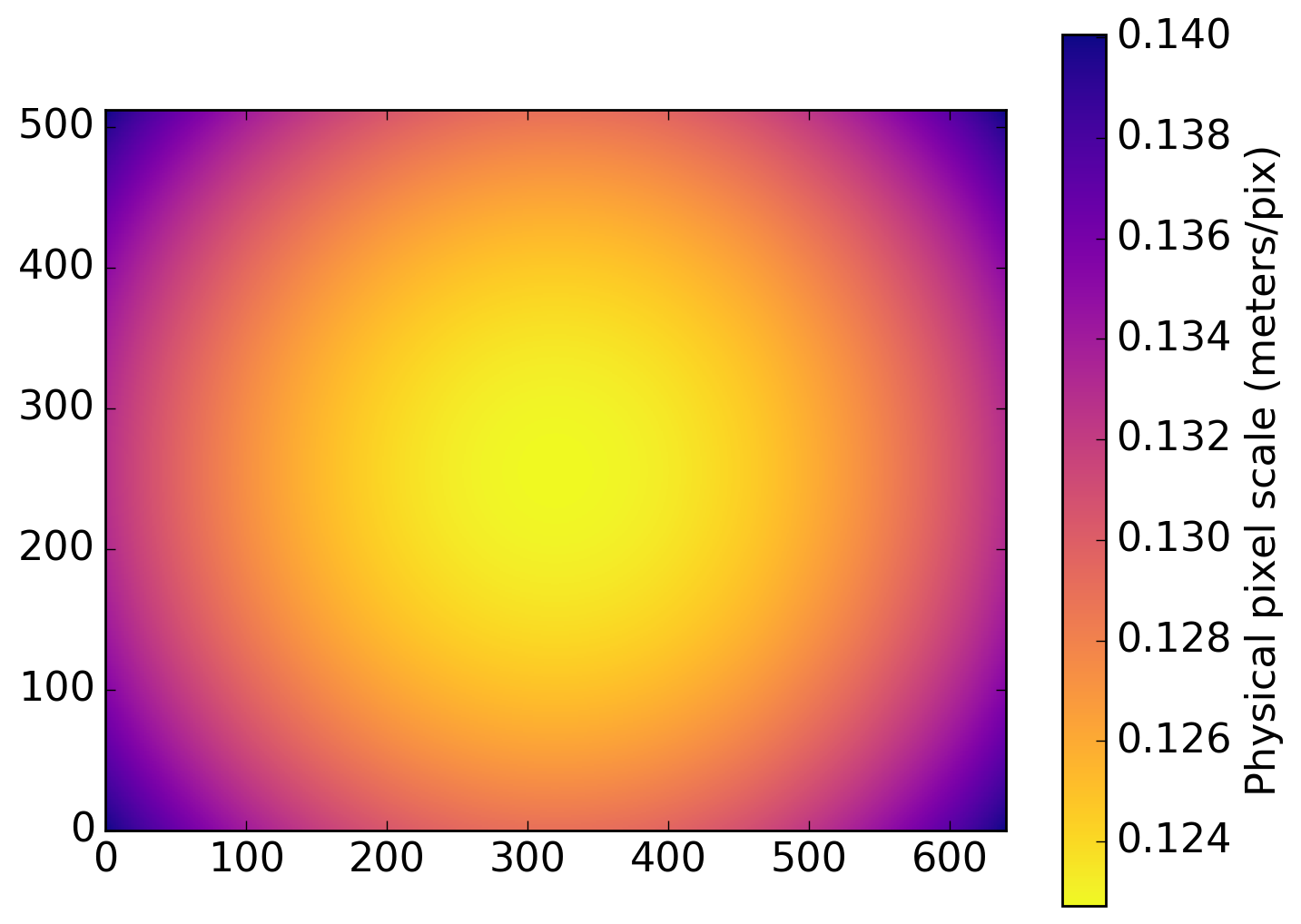}
\includegraphics[width=7.5cm]{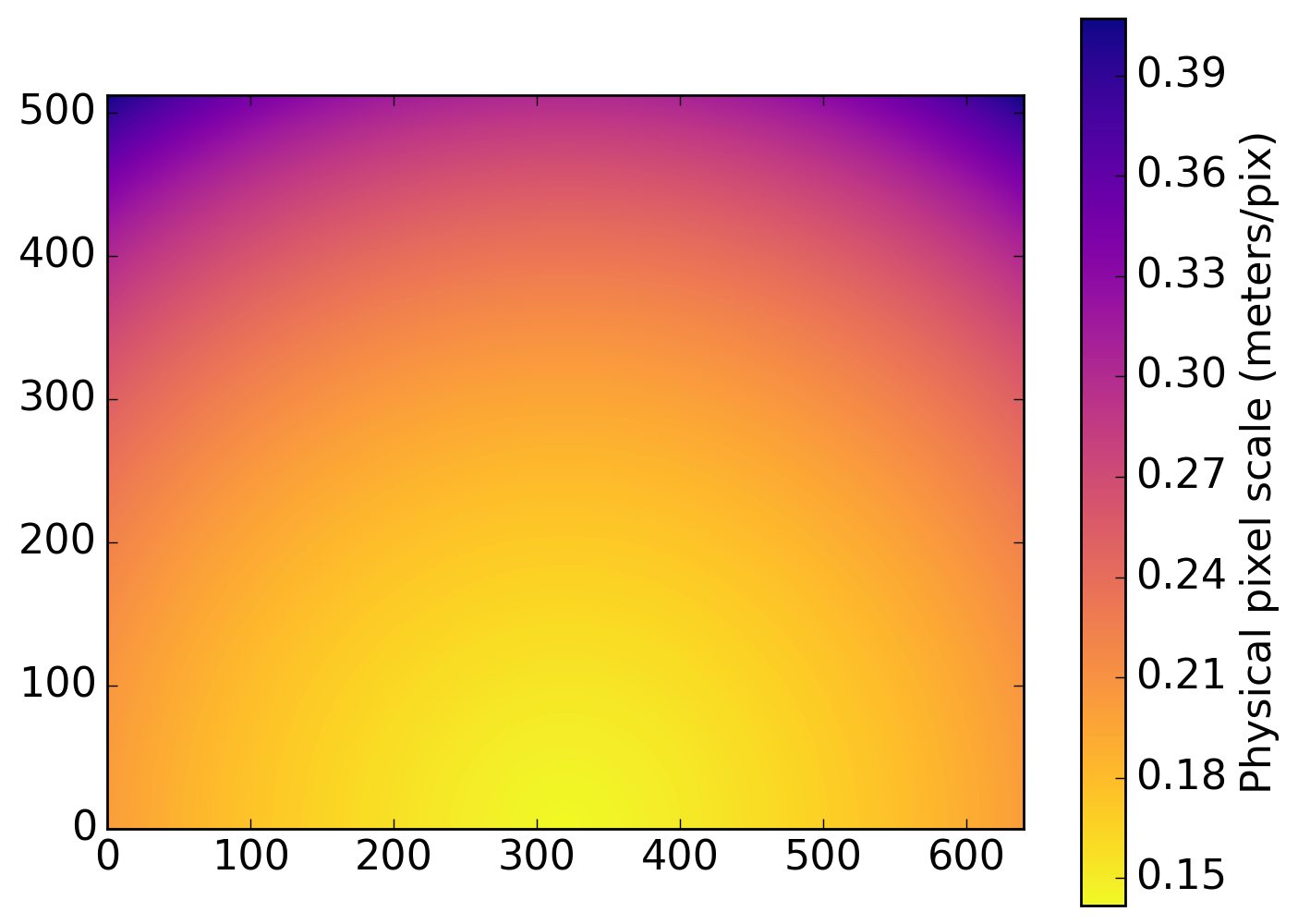}
\includegraphics[width=7.6cm]{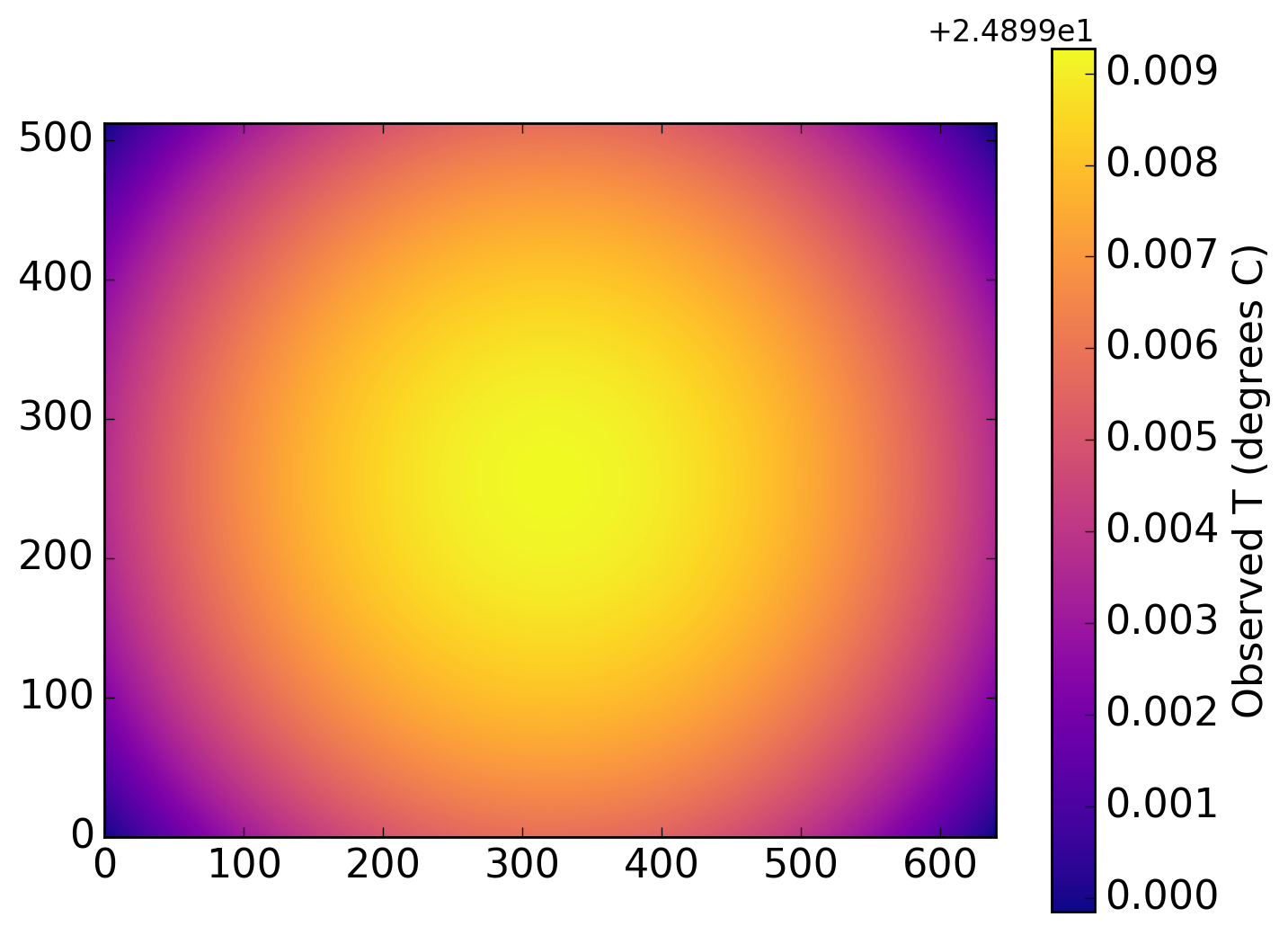}
\includegraphics[width=7.5cm]{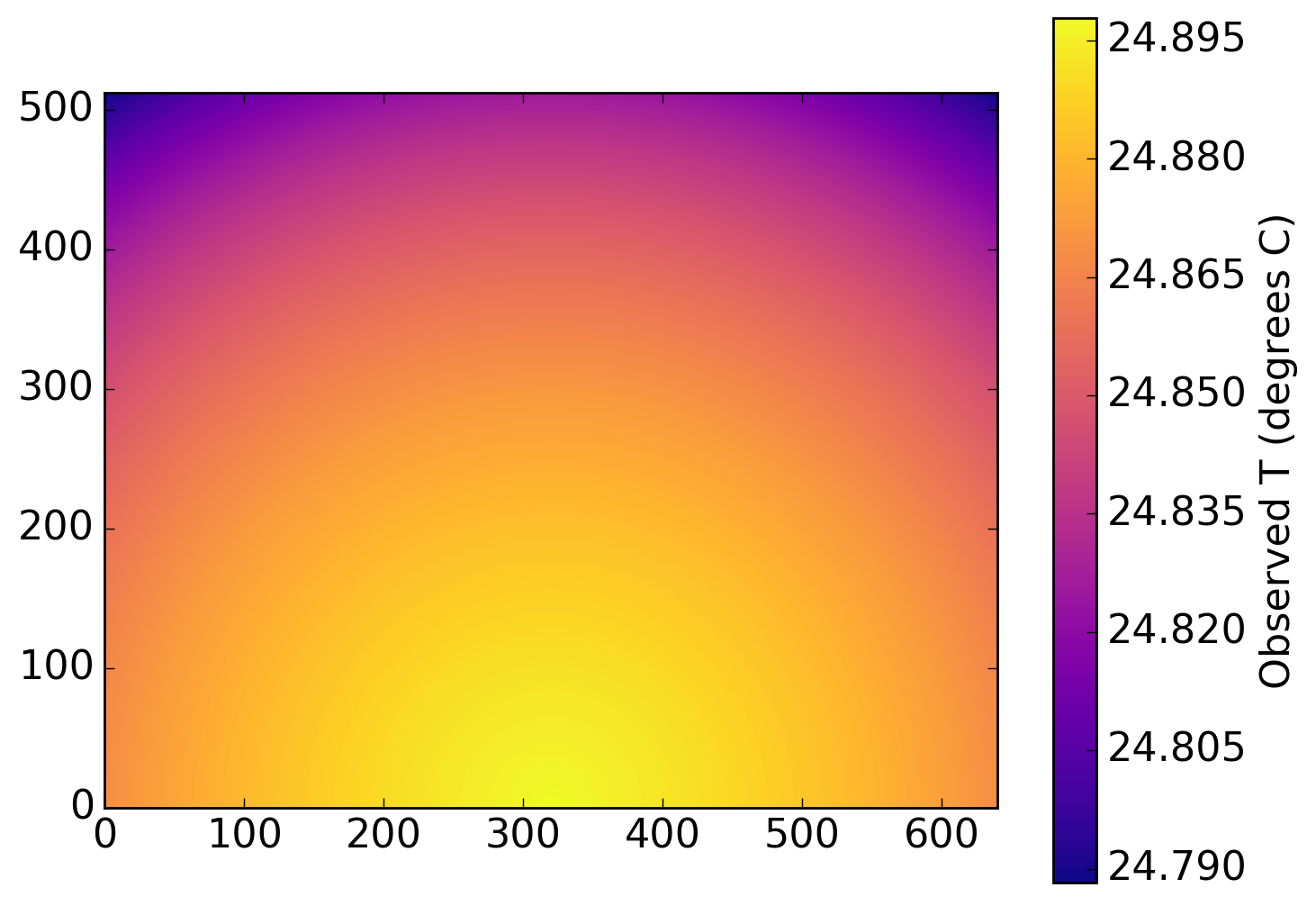}
\caption{Illustration of the observed properties for a TIR camera with $640\times512$ pixels and $45\times37^{\circ}$ field of view, and for drone height of $h=100$ meters with the camera pointed straight down (left column) and with the camera mounted at a $\phi=30^{\circ}$ angle (right column). Top row: distance between camera mounted on drone and point on ground as seen by each pixel. Middle row: Physical pixel scale ($\rho_p$) for each pixel. Bottom row: Observed temperature of a source with actual temperature $T_s=25^{\circ}$C due to absorbed and emitted TIR radiation by the atmosphere, for air temperature of $T_m=20^{\circ}$, pressure of 101kPa, humidity 50\%.}
\label{corrections_arrays}
\end{figure}

This information combined with that presented in Sections~\ref{sec:fov} and Appendix~\ref{app:angle_math} can be used to create a ``corrections array'' which could be applied to data to account for the variation of received temperature with distance due to a non-uniform distance between camera and ground. 
For example, using the camera described in Section~\ref{sec:fov} at a $\phi=30^{\circ}$ angle and height $h=100$ meters, using Equation~\ref{equ:any_R} the closest distance $R_C=116$ meters and the furthest distance $R_F=256$ meters (at the middle of the $x$-axis). 
In this case, two sources with the same actual temperature ($T_s=25^{\circ}$C), in the presence of air of pressure 101kPa, humidity 50\% (as Figures~\ref{fig:spectr_abosorb} and \ref{app_fig:abs_spec}), and air temperature $T_m=10^{\circ}$C, would show a difference in temperature of $\sim-0.1^{\circ}$C between the top and bottom of the field of view (illustrated in right hand side of Figure~\ref{corrections_arrays}). For air temperature much warmer or colder than the source temperature this difference will be much larger. For large values of $h$, in cases where the camera is pointing straight down, the value of $R$ for the edge of the field of view will be larger than that at the centre, as such sources at the edges of the FOV will have differing temperature compared to the same sources if they were viewed in the centre (left hand side of Figure~\ref{corrections_arrays}).

\subsection{The effect of fog}\label{fog_appendix}

Light passing through the atmosphere is scattered by suspended molecules as a result of Rayleigh scattering. The effective area subtended by a scattering particle or molecule is known as the scattering cross section, $\sigma$, and is a function of the wavelength of light being scattered, $\lambda$, and the diameter of the scattering molecule, $d$. The scattering cross section is described by,
\begin{equation}
\sigma = \frac{2\pi^5}{3}\frac{d^6}{\lambda^4}\left(\frac{n^2-1}{n^2+2}\right)^2,
\end{equation}
where $n$ is the refractive index of that molecule \citep[refractive index is also a function of wavelength, see][]{water_ref_index}. The fraction of transmitted, ie the ratio of intensity of light received, $I$, to that originally emitted, $I_0$, as a function of distance $S$ that the light has travelled through the medium is then given by,
\begin{equation}
\frac{I}{I_0}= e^{-N\sigma S},
\end{equation}
where $N$ is the number density of the scattering molecules in the medium (Beer-Lambert law). See \cite{scattering_book} for full details on Rayleigh scattering and the Beer-Lamber law. 

Rayleigh scattering can only occur if $\lambda >>d$. The droplets of water which are suspended in fog have diameters between $1-10\mu$m and thermal infrared light has wavelengths in the range $\lambda = 8-14\mu$m, meaning that Rayleigh scattering is generally applicable here. However, visible light has wavelength $\lambda=0.6-0.9\mu$m which is much smaller than the size of the droplets, so visible light will be refracted by encounters with water droplets \citep[according to Mie theory, see][]{scattering_book}, so to an observer will appear to be strongly scattered.

%
%
%
%


\appendix

\end{document}